\documentclass[aps, prd, twocolumn, showpacs, nofootinbib, floatfix,superscriptaddress]{revtex4-2}
\usepackage{subfigure}
\usepackage{epsfig}
\usepackage{amsmath}
\usepackage{amsfonts}
\usepackage{amssymb}
\usepackage{amssymb,amsmath,amsfonts}
\usepackage{xfrac}
\usepackage{color}
\usepackage[utf8]{inputenc}
\usepackage{graphicx}
\usepackage{dcolumn}
\usepackage{bm}
\usepackage{tikz}
\usepackage{amssymb}
\usepackage{float}
\usepackage{amsmath}
\usepackage{dcolumn}
\usepackage{cancel}

\usepackage{tcolorbox}
\usepackage{hyperref}
\hypersetup{colorlinks, citecolor=red, linkcolor=bluscuro, urlcolor=bluscuro}
\definecolor{rossos}{cmyk}{0,1,1,0.55}
\definecolor{bluscuro}{rgb}{0.15, 0.2, .85}
\definecolor{bluchiaro}{cmyk}{1,.3,0.,0.1}

\newcommand{\bea}{\begin{eqnarray}}
 
\newcommand{\eea}{\end{eqnarray}}
\newcommand{\bma}{\begin{pmatrix}}
\newcommand{\ema}{\end{pmatrix}}
\newcommand{\be}{\begin{equation}}
\newcommand{\ee}{\end{equation}}
\newcommand{\beno}{\begin{equation*}}
\newcommand{\eeno}{\end{equation*}}

\def\doi{http://doi.org}
\def\arxiv{http://arxiv.org/abs}

\renewcommand{\[}{\left[}

\begin{document}

\title{Modified gravity from Weyl connection and the $f(R,\cal{A})$ 
 extension}

\author{Gerasimos Kouniatalis}
\email{gerasimos\_kouniatalis@mail.ntua.gr}
\affiliation{Physics Department, National Technical University of Athens,
15780 Zografou Campus,  Athens, Greece}
 \affiliation{National Observatory of Athens, Lofos Nymfon, 11852 Athens, 
Greece}

\author{Emmanuel N. Saridakis} \email{msaridak@noa.gr}
\affiliation{National Observatory of Athens, Lofos Nymfon, 11852 Athens, Greece}
\affiliation{CAS Key Laboratory for Researches in Galaxies and Cosmology, 
School 
of Astronomy and Space Science, University of Science and Technology of China, 
Hefei, Anhui 230026, China}
\affiliation{Departamento de Matem\'{a}ticas, Universidad Cat\'{o}lica del 
Norte, Avda. Angamos 0610, Casilla 1280 Antofagasta, Chile}


\begin{abstract}

We use  Weyl connection and Weyl geometry in order to construct novel modified 
gravitational theories. In the simplest case where one uses only the  
Weyl-connection  Ricci scalar as a Lagrangian, the theory recovers general 
relativity. However, by upgrading the Weyl field to a dynamical field with a 
general potential and/or general couplings constructed from its trace, leads to 
new modified gravity theories,  where the extra degrees of freedom  arise from 
the Weyl 
field. Additionally, since the Weyl-connection   Ricci scalar differs from the 
 Levi-Civita Ricci scalar by terms up to first derivatives of the Weyl field,  
the resulting field equations for both the metric and the Weyl field are of 
second order, and thus  the theory is free from Ostrogradsky ghosts.   Finally, 
we construct the most general theory, namely the  $f(\tilde{R},\cal{A})$ 
gravity, which is also ghost free. Applying 
the 
above classes of theories at a cosmological framework we  obtain an effective 
dark energy sector of geometrical origin. In the simplest class of theories we 
are able to obtain an effective cosmological constant, and thus we recover 
$\Lambda$CDM paradigm, nevertheless in more general cases we acquire a 
dynamical 
dark energy.
These theories can reproduce the thermal  history of the Universe, and the 
corresponding dark energy equation-of-state parameter presents a 
rich behavior.

\end{abstract}

\maketitle

\section{Introduction}

The concordance  Model of Cosmology, namely   $\Lambda$-Cold Dark Matter 
($\Lambda$CDM) in the framework of general relativity, completed with the 
addition of the inflationary phase, has been proved to be very efficient in 
quantitatively describing the universe behavior 
\cite{Sahni:1999gb,Peebles:2002gy}. Nevertheless, it exhibits some potential 
disadvantages both at the theoretical level, such as the  non-renormalizability 
of general relativity and the cosmological constant problem, as well as at the 
observational level, such as the possibility of an evolving dark energy or 
various tensions between its predictions and particular datasets, such as the  
$H_0$ and $\sigma_8$ tensions \cite{Abdalla:2022yfr}. Hence, in the literature 
one can 
find a huge number of alternatives, that aim to improve its behavior and 
alleviate the tensions. One first direction that one can follow is to maintain
general relativity and alter the content of the universe, namely introduce 
extra particles, field, fluids, or mutual interactions 
\cite{Copeland:2006wr,Cai:2009zp}. The second direction is to    construct
extensions and modifications of general relativity 
\cite{CANTATA:2021ktz,Capozziello:2011et,Cai:2015emx,Nojiri:2017ncd}.

In order to  construct gravitational modifications one can start from the 
standard curvature formulation of gravity and extend the Einstein-Hilbert 
Lagrangian in various ways, resulting to
 $f(R)$ gravity 
\cite{Starobinsky:1980te,Capozziello:2002rd,DeFelice:2010aj,Nojiri:2010wj},
    $f(G)$ gravity \cite{Nojiri:2005jg, DeFelice:2008wz,DeFelice:2009aj},   cubic 
gravity \cite{Asimakis:2022mbe}, 
Lovelock gravity \cite{Lovelock:1971yv, Deruelle:1989fj},  
etc. However, one has equal right to start from the equivalent torsional 
formulation of gravity, namely from the  
Teleparallel Equivalent of General Relativity and extend it in various ways, 
resulting to   $f(T)$ gravity
\cite{Cai:2015emx,Linder:2010py, Chen:2010va}, to $f(T,T_G)$ 
gravity \cite{Kofinas:2014owa,Kofinas:2014daa}, to $f(T,B)$ gravity 
\cite{Bahamonde:2015zma,Bahamonde:2016grb}, etc. Similarly, one can start from 
the equivalent formulation of gravity in terms of non-metricity, and construct
  $f(Q)$ gravity  \cite{BeltranJimenez:2017tkd,Heisenberg:2023lru}, $f(Q,C)$ 
gravity \cite{De:2023xua}, etc. From the above classes of gravitational 
modifications 
one deduces that  the role of the imposed underlying connection is crucial, 
since this leads to different geometrical structure. In particular, in 
curvature gravity ones uses the standard Levi-Civita connection, i.e. 
Riemannian geometry, in torsional gravity one uses the 
Weitzenb$\ddot{\text{o}}$ck connection, i.e. Weitzenb$\ddot{\text{o}}$ck  
geometry, and in non-metric gravity ones uses the symmetric teleparallel 
connection and geometry.

Hermann Weyl introduced a different connection, and thus a different geometry 
quite  early,  incorporating the notion 
of gauge invariance into the structure of spacetime geometry itself  
\cite{Weyl:1918ib}. Starting from the Weyl transformations, which  do not 
preserve the form  of the covariant derivative, Weyl introduced a  gauge field 
$A_{\mu}$ in the connection to restore consistency, and this  Weyl 
connection gives rise to Weyl geometry \cite{Weylbook}. We mention here that 
  Weyl transformations and conformal  transformations do not coincide, since in 
the latter case   the conformal factor is a specific function associated with a 
diffeomorphism that is a conformal symmetry of the theory, whereas in Weyl 
transformations it is arbitrary.

In this work we are interesting in constructing gravitational theories based 
on Weyl connection and geometry. Note that although
Weyl's original motivation was a unified description of gravity and 
electromagnetism in geometrical terms, this did not prove to be the case, 
however the richer geometrical structure offers us the motivation to 
use it in order to construct richer gravitational theories (interestingly 
enough the 
unification of gravity and  electromagnetism in geometrical terms was also 
Einstein's motivation to include torsion, which did   not work either but 
offered us richer geometries to construct gravitational modifications).
Additionally, apart from the basic scenario, 
we will extend them by introducing functions of the trace of the  Weyl   
field in the action.

We stress here that the theories that are   going to be 
constructed, namely modified gravity through Weyl connection, should not be 
confused with ``Weyl gravity'' which is a curvature-based modified 
gravity that uses the Weyl tensor 
\cite{Mannheim:1988dj,Zee:1981ff,Zee:1983mj,Kazanas:1988qa} and its 
cosmological and black hole
applications 
\cite{Sola:1988nz,Mannheim:1990ya,La:1991nk,Elizondo:1994vh,Bronnikov:1997gj,
Edery:1997hu,Klemm:1998kf,Boulanger:2001he,Pireaux:2004xb,Pireaux:2004id,
Flanagan:2006ra,Edery:2006hg,Lobo:2008zu,Sultana:2010zz,Percacci:2011uf,
Dengiz:2011ig,Tanhayi:2012nn,Sultana:2012qp,Deruelle:2012xv,Said:2012xt,
Cattani:2013dla, 
Wheeler:2013ora,Quiros:2014hua,Myung:2014jha,Cusin:2015rex,Mureika:2016efo, 
Oda:2016psn,Ghilencea:2018dqd,Zinhailo:2018ska,
Ghilencea:2018thl,Ghilencea:2019jux, 
Ghilencea:2020rxc,Takizawa:2020dja,Jawad:2020wlg,Geiller:2021vpg,Yang:2022icz, 
Hell:2023rbf,
Karananas:2021gco,Roumelioti:2024lvn,Gialamas:2024iyu}.  
Additionally, the presented theories are   radically different and   more 
general from the use of Weyl field within the symmetric teleparallel framework, 
namely in the $f(Q,T)$ theories
\cite{Haghani:2012bt,Haghani:2013pea,Haghani:2014zra,Xu:2020yeg,Yang:2021fjy,
Gadbail:2021kgd,Harko:2021tav,
Berezin:2021iof,Gadbail:2021fjf,Berezin:2022phu,Berezin:2022odj,
Gadbail:2022scf,Koussour:2023nqr,Gadbail:2023enu,Bhardwaj:2023lph,
Koussour:2024aez,Zhadyranova:2024hbc,Sakti:2024pze,Banados:2024rfy,
Harko:2024fnt,Bhagat:2024nac}, or from its use in 
quadratic vector-tensor theories 
\cite{BeltranJimenez:2016wxw,BeltranJimenez:2015pnp}.
As we will see, the richer 
geometrical structure of Weyl connection and geometry, when applied in a 
cosmological framework, will give rise to richer and interesting cosmological 
phenomenology.

The plan of the work is the following: In Section \ref{model} we review  Weyl 
connection and Weyl geometry, and then we construct various 
classes of gravitational modifications based on it. Then in Section 
\ref{Cosmology} we apply these theories at a cosmological framework, 
extracting the Friedmann equations and presenting a specific example. Finally, 
Section \ref{Conclusions} is devoted to the conclusions.

\section{Modified  gravity from  Weyl geometry}
\label{model}

In this section we briefly review the Weyl connection and Weyl geometry and 
then we use it to construct actions for gravitational theories.

  \subsection{  Weyl connection and geometry}

Let us start with the basics of  Weyl connection and   geometry. For the moment 
we remain in $d$ dimensions and later on we will focus on the $d=4$ case.
Weyl transformations are defined by 
\cite{Weyl:1918pdp,Romero:2012hs,Wheeler:2018rjb}
\begin{equation} \label{metrictransform}
    g \rightarrow \mathcal{B}^{-2}(x) g,
\end{equation}
where  $\mathcal{B}(x)$ is a completely arbitrary function of spacetime 
coordinates, in contrast to the conformal transformations $ g \rightarrow 
\omega(x)^{-2}g$ where $\omega(x)$ is  associated to   conformal symmetry.
 Since Weyl transformations   do not preserve the form of the 
Levi-Civita covariant derivative, i.e.  $  
    \nabla g \rightarrow \nabla(\mathcal{B}(x)^{-2}g) 
    =
    \left( \nabla g - 2gd\ln{\mathcal{B}(x)}  \right) \mathcal{B}(x)^{-2} $,
Weyl introduced a gauge 
field $A_{\mu}$ which transforms as 
\begin{equation}\label{Weyltr2}
    A_{\mu} \rightarrow A_{\mu} - \partial_{\mu}\ln{\mathcal{B}(x)},
\end{equation}
and then he introduced a Weyl-invariant connection as
\begin{equation} \label{Weyl connection}
    \tilde{\Gamma}^{\lambda}_{\mu\nu} \equiv \Gamma^{\lambda}_{\mu\nu}  - 
(A_{\mu} \delta^{\lambda}{}_{\nu} + A_{\nu} \delta^{\lambda}{}_{\mu} - 
A^{\lambda}g_{\mu\nu}  ) ,
\end{equation}
where $    \Gamma^{\lambda}_{\mu\nu} = \frac{1}{2} 
g^{\lambda\rho}(\partial_{\mu}g_{\rho\nu} + \partial_{\nu}g_{\rho\mu} - 
\partial_{\rho}g_{\mu\nu}) $ is the Levi-Civita  connection. Note that 
 the introduction of the Weyl   field does not destroy the symmetry, i.e. 
$\Tilde{\Gamma}^{\lambda}_{\mu\nu} = \tilde{\Gamma}^{\lambda}_{\nu\mu}$ and 
thus Weyl connection   has zero torsion, too. Nevertheless it does have 
non-metricity, since
\begin{equation}
    {\tilde{\nabla}}_{\mu}g_{\alpha\beta} = 2A_{\mu}g_{\alpha\beta},
\end{equation}
with $ {\tilde{\nabla}}_{\mu}$ the covariant derivative corresponding to the 
Weyl connection, whose action on a vector $ X_{\nu}$ is defined as   $  
{\tilde{\nabla}}_{\mu} X_{\nu} \equiv \partial_{\mu} X_{\nu} - 
  \tilde{\Gamma}^{\lambda}_{\mu\nu} X_{\lambda}$  (and thus
  $\tilde{\nabla}_\mu X^\nu = \nabla_\mu X^\nu - A_{\mu}X^{\nu} - A_{\lambda} 
X^{\lambda} \delta^{\nu}{}_{\mu} + X_{\mu}A^{\nu}$). Hence,  the Weyl 
covariant derivative of the metric tensor is proportional to 
the metric itself, scaled by the Weyl vector.

The Riemann tensor corresponding to the Weyl connection reads as
\begin{equation}
    \Tilde{R}^{\lambda}{}_{\mu\rho\nu} = \partial_{\rho} 
\Tilde{\Gamma}^{\lambda}_{\mu\nu} - \partial_{\nu} 
\Tilde{\Gamma}^{\lambda}_{\mu\rho} + 
\Tilde{\Gamma}^{\kappa}_{\mu\nu} \Tilde{\Gamma}^{\lambda}_{\kappa\rho}
    -  \Tilde{\Gamma}^{\kappa}_{\mu\rho} \Tilde{\Gamma}^{\lambda}_{\kappa\nu},
\end{equation}
 while the Ricci tensor and the Ricci scalar are respectively given by
$
        \Tilde{R}_{\mu\nu} = \delta^{\rho}{}_{\lambda} 
\Tilde{R}^{\lambda}{}_{\mu\rho\nu}$ and $ \Tilde{R} = g^{\mu\nu} 
\Tilde{R}_{\mu\nu}$. 
As one can see, under Weyl transformations we have 
\begin{equation} \label{Transformations of Ricci}
\begin{aligned}
    \Tilde{R}^{\lambda}{}_{\mu\rho\nu} &\rightarrow 
\Tilde{R}^{\lambda}{}_{\mu\rho\nu} \\[1em]
    \Tilde{R}_{\mu\nu} &\rightarrow \Tilde{R}_{\mu\nu} \\[1em]
    \Tilde{R} &\rightarrow \mathcal{B}(x)^{2} \Tilde{R} ,
\end{aligned}
\end{equation}
namely  the Riemann and Ricci tensors are Weyl-invariant, or equivalently they 
have  zero Weyl-weight, while  the Ricci scalar is covariant    under Weyl 
transformations, or equivalently it has 
Weyl-weight equal $2$ (if a tensor $X$ under Weyl transformations 
(\ref{metrictransform}) and (\ref{Weyltr2}) transforms as 
$
    X \rightarrow \mathcal{B}(x)^{w} X$ then its  Weyl-weight is $w$).
Nevertheless, note that although  the Riemann tensor 
is   antisymmetric in the last two indices (as the Riemann tensor corresponding 
to the Levi-Civita connection), it 
lacks the antisymmetry of the first two indices and the interchange symmetry of 
the index pairs. Concerning the Ricci tensor, it has an antisymmetric part, 
which is 
\begin{equation} \label{Antisymmetry of Ricci}
    \Tilde{R}_{[\mu\nu]} = -d  F_{\mu\nu},
\end{equation}
(we use the antisymmetry notation as $\tilde{R}_{[\mu\nu]} \equiv 
\tilde{R}_{\mu\nu}-\tilde{R}_{\nu\mu}$) where $d$ is the number of dimensions 
and $F_{\mu\nu}$ is the 
field-strength tensor of $A_{\mu}$, defined by 
\begin{equation} \label{Field-strength}
 F_{\mu\nu} = \tilde{\nabla}_{\mu} A_{\nu} - 
\tilde{\nabla}_{\nu}A_{\mu}. 
\end{equation}

Note that  although we defined the field-strength tensor using the Weyl 
covariant derivative, we could use the  Levi-Civita covariant derivative, or 
even the partial derivative, since $\tilde{\nabla}_{\mu} A_{\nu} - 
\tilde{\nabla}_{\nu}A_{\mu}=  \nabla_{\mu}A_{\nu} - 
\nabla_{\nu}A_{\mu} =  
\partial_{\mu}A_{\nu} - \partial_{\nu}A_{\mu} ,
$
since   both the Levi-Civita and Weyl connections are torsion-free.
Finally, note that the  field-strength tensor of $A_{\mu}$ 
  is   Weyl-invariant, since $
    F_{\mu\nu} = \nabla_{[\mu}A_{\nu]} = \partial_{[\mu}A_{\nu]}  
\rightarrow \partial_{[\mu}A_{\nu]}  = F_{\mu\nu}$.
All the above can be easily seen if we express the Weyl-connection quantities 
in terms of the Levi-Civita ones, namely 
\begin{eqnarray}
&&      
\!\!\!\!\!
\Tilde{R}^{\lambda}{}_{\mu\rho\nu} = R^{\lambda}{}_{\mu\rho\nu} 
    + \delta^{\lambda}{}_{\mu} \nabla_{[\nu}A_{\rho]} 
    + \delta^{\lambda}{}_{[\rho}\nabla_{\nu]}A_{\mu} 
    + g_{\mu[\nu}\nabla_{\rho]}A^{\lambda} \nonumber \\ 
    &&
    \ \ \ \ \ \ \ \ \
    + (A_{\mu}A_{[\nu} - A^2g_{\mu[\nu})\delta^{\lambda}{}_{\rho]}
    + A^{\lambda}A_{[\rho}g_{\nu]\mu} \\ 
        && \!\!\!\!\!\Tilde{R}_{\mu\nu} = R_{\mu\nu}
    -\frac{d}{2} F_{\mu\nu} +  \nabla_{(\mu}A_{\nu)} 
    + \nabla_{\lambda} A^{\lambda} g_{\mu\nu}\nonumber \\ 
&    &    \ \ \ \ \ \ \ + 
    (d-2) (A_{\mu}A_{\nu}- A_{\lambda}A^{\lambda} g_{\mu\nu}) \\
    && \!\!\!\!\!
    \Tilde{R} =   R + 2(d-1) \nabla^\nu   A_\nu  - (d-1)(d-2)A_\mu A^\mu.
    \label{WeylRicciscalar}
 \end{eqnarray}
  These relations will be   useful later, when we 
will consider the modified version of the Einstein-Hilbert action. Finally,
note that we can always recover the the standard Levi-Civita Riemann tensor, 
Ricci tensor and Ricci scalar, by just choosing (i.e. gauge-fixing) $A_{\mu} =0 
$, and thus Riemannian geometry is a special case of Weyl geometry. 

We mention here that in principle, as all non-metricity theories, Weyl geometry 
exhibits   non-integrability of vector lengths, however this can be 
  addressed by considering Weyl integrable spaces, where the Weyl 
vector is derived from a scalar field $A_\mu = \partial_\mu \phi$, allowing for 
a coherent integration of lengths along closed paths and making the theory more 
compatible with physical observations \cite{Scholz:2017pfo}.

  \subsection{ Gravity on Weyl geometry  }
  
  In the previous subsection we presented the basics of  Weyl connection and 
geometry. In this section we proceed by constructing a gravitational theory on 
Weyl geometry. Without loss of generality, form now on we focus on $d=4$ 
dimensions.

\subsubsection{Class I}

Following the standard lore, a simplest choice would be 
  \begin{equation}
    S   =  \frac{1}{16\pi G}
    \int d^4x \sqrt{-g}  \Tilde{R} +S_m,
    \label{action0}
\end{equation}
where   $\Tilde{R}$ is the Ricci scalar corresponding to the Weyl connection 
given in (\ref{WeylRicciscalar}),
and $S_m$ is the matter action.  
 Performing variation with respect to  the metric   
we obtain 
\begin{equation}
     R^{\mu}{}_{\nu} - \frac{1}{2} 
R\delta^{\mu}{}_{\nu} + 3A^2 \delta^{\mu}{}_{\nu} - 6A^{\mu}A_{\nu}  = 8\pi G 
T^{\mu}{}_{\nu},
    \label{Weylfieldeqs}
\end{equation}
where 
 $T^{\mu}{}_{\nu}$ is the usual energy-momentum 
tensor corresponding to the matter action (assuming that the matter Lagrangian 
depends only on the metric and the matter fields, and not on $A_{\mu}$).
Furthermore, since inside $ \Tilde{R}$ we also have the
Weyl gauge field,   we must additionally perform variation of (\ref{action0}) 
in 
terms of $A^{\mu}$, finally obtaining 
\begin{equation}
A^{\mu} = 0,
\end{equation}
which is just a constraint as expected, since no dynamical terms were included 
in  (\ref{action0}). Nevertheless, this trivial constraint implies that Weyl 
geometry recovers Riemannian geometry, and the above simple action recovers 
general 
relativity, with no new information being gained.

\subsubsection{Class II}

Having the above in mind, we proceed by upgrading the Weyl field to a 
dynamical one, and we include in the 
Lagrangian its kinetic term, namely   $ - \frac{1}{4} 
F_{\mu\nu}F^{\mu\nu} $. Additionally, using $A_{\mu}$ we can 
immediately construct a new scalar, namely its trace  ${\cal{A}}\equiv 
A_{\mu}A^{\mu}$, and then 
extend the 
Lagrangian by considering arbitrary functions  $f(\cal{A})$ (and thus this 
class of theories is radically different than those examined in 
\cite{Haghani:2012bt,Haghani:2013pea,Haghani:2014zra,Xu:2020yeg,Yang:2021fjy,
Gadbail:2021kgd,Harko:2021tav,
Berezin:2021iof,Gadbail:2021fjf,Berezin:2022phu,Berezin:2022odj,
Gadbail:2022scf,Koussour:2023nqr,Gadbail:2023enu,Bhardwaj:2023lph,
Koussour:2024aez,Zhadyranova:2024hbc,Sakti:2024pze,Banados:2024rfy,
Harko:2024fnt,Bhagat:2024nac,BeltranJimenez:2016wxw,BeltranJimenez:2015pnp}). 
Hence, a 
modified 
gravitational action built on  Weyl geometry  would be:
 \begin{equation}
    S =  \frac{1}{16\pi G}
    \int d^4x \sqrt{-g} \left[ \Tilde{R}+  f({\cal{A}}) - \frac{1}{4} 
F_{\mu\nu}F^{\mu\nu} \right]+S_m.
\label{fullaction}
\end{equation}
 Performing variation with respect to  the metric and using 
(\ref{WeylRicciscalar}) for $d=4$
yields
\begin{equation}
    R^{\mu}{}_{\nu} - \frac{1}{2} 
R\delta^{\mu}{}_{\nu}  + K^{\mu}{}_{\nu}  = 8\pi G T^{\mu}{}_{\nu},
    \label{Weylfieldeqs}
\end{equation}
where  
  we have defined the tensor
\begin{eqnarray}
 &&\!\!\!\!\!\!\!\!\!\!\!\!\!\!\!\!\!\!\!
     K^{\mu}{}_{\nu} =  
      \left[3{\cal{A}} -  
\frac{1}{2} f({\cal{A}}) + \frac{1}{4} F_{\alpha\beta}\nabla^{\alpha}A^{\beta} 
\right] \delta^{\mu}{}_{\nu}
   \nonumber\\
   &&
    +  \left[ f'({\cal{A}})
    - 6 \right] A^{\mu}A_{\nu},
    \label{KtensorclassII}
 \end{eqnarray}
 with the prime denoting derivative with respect to ${\cal{A}}$.
 Additionally, varying the action (\ref{fullaction}) with respect to  
$A_{\mu}$, and using (\ref{WeylRicciscalar}) for $d=4$, we obtain 
\begin{equation}
        \tilde{\nabla}_{\alpha}F^{\alpha\mu} 
        +4A_{\alpha}F^{\alpha\mu} + 2A^{\mu}[f'(\mathcal{A})-6] = 0,
\end{equation} 
which can be expressed in terms of the standard Levi-Civita covariant 
derivative as
  \begin{equation}
\nabla_{\alpha} F^{\alpha\mu} 
+  2  f'({\cal{A}}) A^{\mu} 
-  12A^{\mu}
     =0 .
     \label{Weylconnectioneqs}
\end{equation}
 Finally, note that imposing the matter conservation equation  $\nabla_{\mu} 
T^{\mu}{}_{\nu} = 0$, equation  (\ref{Weylfieldeqs}) implies 
\begin{equation} \label{ConservK}
    \nabla_{\mu} K^{\mu}{}_{\nu} = 0,
\end{equation}
which is the conservation equation corresponding to the Weyl field.

In summary, the modified gravity theory of this class, constructed using the 
Weyl 
connection, is not trivial and indeed exhibits a richer structure. 
As one can see it has one  extra vector degree of 
freedom comparing to general relativity. 
{In particular,  the theory   has the extra Weyl gauge field $A_{\mu}$, 
which originally  contributes four degrees of 
freedom, however, due to the Weyl-integrability condition $A_{\mu} = 
\partial_{\mu}\phi$, one degree of freedom is eliminated, resulting in a total 
of three additional propagating modes beyond those of general relativity. 
Hence, these classes of theories have the same number of degrees of freedom 
with Type 3 types of New General Relativity studied in 
 \cite{Tomonari:2024ybs,Tomonari:2024lpv} (see also 
 \cite{Gialamas:2024uar,Capozziello:2024lsz,Barriga:2024hpe}).
}

As expected, due to the fact that action (\ref{fullaction}) is linear in the 
Weyl-connection   Ricci scalar, which differs from the Levi-Civita Ricci scalar 
by at most first derivatives of the Weyl field, the resulting field 
equations are not higher-order, and thus the theory is free from Ostrogradsky 
ghosts \cite{Ostrogradsky}, while for the same reason the  Weyl field equation 
of motion does not 
contain  higher-order terms too. Hence, the theory at hand is ghost free. 
 Definitely,  we notice that in the case $A_{\mu}=0$ equation 
(\ref{Weylfieldeqs})  recovers the standard Einstein field equations, while the 
connection equation (\ref{Weylconnectioneqs}) disappears.

\subsubsection{Class III}

One can consider a further extension of the above class, by considering  of 
theories arising from the gravitational action   
 \begin{equation}
 \begin{aligned}
         S
    =  \frac{1}{16\pi G} \int d^4x \sqrt{-g} 
    &\left[ \tilde{R} +  
h(\mathcal{A})A^{\mu}A^{\nu}\tilde{\nabla}_{\mu}A_{\nu} 
\right.\\
    &\left.+f(\mathcal{A})- \frac{1}{4} 
F_{\mu\nu}F^{\mu\nu} \right],
\label{actionclassIII}
 \end{aligned}
\end{equation}
where $h(\mathcal{A})$ is another arbitrary function of $\mathcal{A}$.
  Varying the total action $S+S_m$ with respect to the metric    we obtain the 
field equations 
(\ref{Weylfieldeqs}), but now
{\small{
\begin{eqnarray}
&&     \!\!\!\!     K^{\mu}{}_{\nu}=
 \frac{1}{2}F_{\nu\alpha}F^{\alpha\mu}
        + \frac{1}{2}h(\mathcal{A})A^{\alpha}A^{(\mu}\nabla_{\nu)}A_{\alpha}
  \nonumber       \\
        && \!\!\!\!  +  \left[3 {\mathcal{A}} 
       -\frac{h({\mathcal{A}} )}{2} 
        \left({\mathcal{A}} ^2
      +
A^{\alpha}A^{\beta}\nabla_{\beta}A_{\alpha}\right)
      +\frac{1}{4}\nabla^{\beta}A^{\alpha}F_{\beta \alpha} 
      -\frac{f(\mathcal{A})}{2} \right]\!\delta^{\mu}{}_{\nu}   \nonumber
 \\
        && \!\!\!\!  
        + \frac{1}{2}A^{\mu}A_{\nu}\Big\{2f'({\mathcal{A}} 
)-12-h({\mathcal{A}} )\nabla_{\lambda} A^{\lambda}   \nonumber    \\
        && \ \ \ \ \ \ \ \ \ \   + 2{\mathcal{A}} [2h({\mathcal{A}} 
)+{\mathcal{A}} h'({\mathcal{A}} )]\Big\},
        \label{KtensorclassIII}
  \end{eqnarray}}}
while variation  with respect to the Weyl   field yields
 \begin{eqnarray}
&& \! \! \! \! \! \! \! \! \!\nabla_{\alpha} F^{\alpha\mu} 
+h(\mathcal{A})A^{\alpha}\nabla^{\mu}A_{\alpha} 
+A^{\mu}\Big\{2f'(\mathcal{A})-h(\mathcal{A})\nabla_{\lambda} A^{\lambda} 
\nonumber\\
&&
\ \ \ \ \ \ 
-12+ 
2\mathcal{A}\left[2h(\mathcal{A})+A^2h'(\mathcal{A})\right]\Big\}
     =0 
          \label{Weylconnectioneqs22},
\end{eqnarray}
expressed in terms of the usual Levi-Civita connection.
Imposing the matter conservation equation  $\nabla_{\mu} 
T^{\mu}{}_{\nu} = 0$, equation  (\ref{Weylfieldeqs}) implies 
\begin{equation} \label{ConservKIII}
    \nabla_{\mu} K^{\mu}{}_{\nu} = 0,
\end{equation}
  too.  
  
  {
In summary, similarly to the previous case, this class of theories also has  
 three additional propagating modes beyond those of general relativity.
}
As expected, due to the fact that action (\ref{actionclassIII}) is 
linear in 
the 
Weyl-connection   Ricci scalar, and does not contain more than first 
derivatives of the Weyl field, the resulting field equations for both the 
metric and the Weyl field are of second order, and thus  the theory is free 
from 
Ostrogradsky ghosts.  
 
{
We close this subsection with the following comment.  The classes of   theories 
built up to now, fall within the generalized Proca class, specifically 
within the $\mathcal{L}_3$ subclass of \cite{Heisenberg:2014rta}. This 
correspondence arises since our theory includes 
an additional vector field $A_{\mu}$, 
stemming from the Weyl connection, which, when promoted to a dynamical field, 
exhibits self-interactions and couplings that appear 
in generalized Proca theories too. 
However, in our case the vector field   is not introduced ad hoc, but rather 
emerges naturally from the underlying Weyl connection. Hence, in this 
framework   $A_{\mu}$ 
originates from the non-metricity geometrical property of the connection, not 
from an independent   
field added to the action,   which may act as an advantage.    }

\subsubsection{Class IV}
  
Finally, one can extend the above action to the most general class, namely 
   \begin{equation} \label{GeneralAction}
    \begin{aligned}
         S =  \frac{1}{16\pi G}
    \int d^4x \sqrt{-g} \left[ \Tilde{R}+  f(\Tilde{R},\mathcal{A}) - 
\frac{1}{4} 
F_{\mu\nu}F^{\mu\nu} \right]. 
    \end{aligned}
\end{equation}
In this case, variation of the total action $S+S_m$ in terms of the metric 
yields the  field 
equations   
(\ref{Weylfieldeqs}),  
but now the tensor $    K_{\mu\nu} $ is given by
 \begin{eqnarray} \label{GeneralK}
   &&     K_{\mu\nu} = 3\mathcal{A} g_{\mu\nu} 
        - \frac{1}{2}fg_{\mu\nu} 
        + \frac{1}{4} g_{\mu\nu}F_{\alpha\beta}\nabla^{\alpha\beta} 
        + \frac{1}{2} F_{\mu\alpha}F^{\alpha}{}_{\nu} \nonumber \\
        && +(R_{\mu\nu} 
        + 3g_{\mu\nu}\nabla_{\beta}A^{\beta}) f_{\tilde{R}}    
+A_{\mu}A_{\nu}f_A  -6A_{\mu}A_{\nu}(1+f_{\tilde{R}})
    \nonumber \\
        &&     + 
        \Big[
        6 
A^{\alpha}A^{\beta}g_{\mu\nu}\nabla_{\beta}A_{\alpha}  
   +2A^{\alpha}g_{\mu\nu}\nabla^2A_{\alpha}      
+2g_{\mu\nu}\nabla_{\beta}A_{\alpha}\nabla^{\beta}A^{\alpha} 
\nonumber \\
        &&  
 -6A^{\alpha}A_{\nu}\nabla_{\mu}A_{\alpha} 
        -2\nabla_{\mu}A^{\alpha}\nabla_{\nu}A_{\alpha}      \nonumber \\
        && 
     -2A^{\alpha}\nabla_{\nu}\nabla_{\mu}A_{\alpha}
             -6A^{\alpha}A_{\mu}  
\nabla_{\nu}A_{\alpha}  
     \Big]
f_{\tilde{R}\mathcal{A}} 
\nonumber \\        
&&    +   
        4A_{\mu\nu} f_{\tilde{R}\mathcal{A}\mathcal{A}}  
-36\nabla_{\alpha}\nabla_{\mu}A^{\alpha}\nabla_{\beta}\nabla_{\nu}A^{\beta}  
      \nonumber \\        
&&     +
          \Big[
          6A^{\alpha}\nabla_{\nu}R_{\mu\alpha}    
+18A^{\alpha}A_{\nu}R_{\mu\alpha}  
        +3A^{\alpha}g_{\mu\nu}\nabla_{\alpha}R    
+g_{\mu\nu}\nabla^2R  
\nonumber \\
&&
-18A_{\nu}\nabla_{\alpha}\nabla_{\mu}A^{\alpha} 
-6\nabla_{\alpha}\nabla_{\nu}\nabla_{\mu}A^{\alpha}  
-36A^{\alpha}A^{\beta}g_{\mu\nu}\nabla_{\beta}A_{\alpha}  
\nonumber \\
        && 
+18A^{\alpha}g_{\mu\nu}\nabla_{\beta}\nabla_{\alpha}A^{\beta} 
 -12A^{\alpha}g_{\mu\nu}\nabla^2A_{\alpha}  
        +6g_{\mu\nu}\nabla^2(\nabla_{\beta}A^{\beta}) 
    \nonumber \\
        &&    
-6R_{\mu\beta\nu\alpha}\nabla^{\beta}A^{\alpha} 
-12g_{\mu\nu}\nabla_{\beta}A_{\alpha}\nabla^{\beta}A^{\alpha}  
+36A^{\alpha}A_{\nu}\nabla_{\mu}A_{\alpha} 
           \nonumber \\
        &&  +6R_{\nu\alpha}\nabla_{\mu}A^{\alpha}    -3A_{\nu}\nabla_{\mu}R   
+12\nabla_{\mu}A^{\alpha}\nabla_{\nu}A_{\alpha}  
+6R_{\mu\alpha}\nabla_{\nu}A^{\alpha}  
   \nonumber \\
        &&    
+12A^{\alpha}\nabla_{\nu}\nabla_{\mu}A_{\alpha} 
      -\nabla_{\nu}\nabla_{\mu}R -3A_{\mu}  
(6\nabla_{\alpha}\nabla_{\nu}A^{\alpha}+\nabla_{\nu}R)
\nonumber \\
&&
-18A^{\alpha}A^{\beta}g_{\mu\nu}R_{\alpha\beta}    
   +18A^{\alpha}A_{\mu} R_{\nu\alpha} 
  +36 A^{\alpha}A_{\mu}  
\nabla_{\nu}A_{\alpha}    
      \Big]
      f_{\tilde{R}\tilde{R}}
      \nonumber\\
      &&
      + 
\Big[
24(g_{\mu\nu}A^{\alpha}\nabla^{\beta}A_{\alpha}\nabla_{\gamma}\nabla_{\beta
}A^{ \gamma} - 
A^{\alpha}A^{\beta}g_{\mu\nu}R_{\beta\gamma}\nabla^{\gamma}A_{\alpha})  
\nonumber \\
        &&     -12A^{\alpha}\nabla_{\beta}\nabla_{(\nu}A^{\beta}\nabla_{\mu)}A_{
\alpha} 
+4A^{\alpha}g_{\mu\nu}\nabla_{\beta}R\nabla^{\beta}A_{\alpha} 
      \nonumber \\
&&  
-48(A^{\alpha}A^{\beta}g_{\mu\nu}\nabla_{\gamma}A_{\beta}\nabla^{\gamma}A_{
\alpha}-A^{\alpha}A^{\beta}\nabla_{\mu}A_{\alpha}\nabla_{\nu}A_{\beta}) 
\nonumber \\
&&
+12A^{\alpha}A^{\beta}R_{\beta(\nu}\nabla_{
\mu)}A_{\alpha}    
-2A^{\alpha}\nabla_{\mu}R\nabla_{\nu}A_{\alpha}
\Big]
f_{\tilde{R}\tilde{R}\mathcal{A}}    \nonumber \\
&& 
+\Big\{
36A^{\alpha}R_{\alpha(\nu}\nabla_{\beta}\nabla_{\mu)}A^{\beta} 
-36A^{\alpha}A^{\beta}R_{\mu\alpha}R_{\nu\beta}  
 \nonumber \\
        &&+g_{\mu\nu}(\nabla R)^2 
+12g_{\mu\nu}\nabla^{\alpha}R\nabla_{\beta}\nabla_{\alpha}A^{\beta}   \nonumber 
\\
&&+144[(A^{\alpha}A^{\beta}g_{\mu\nu}\nabla_{\gamma}A_{\beta}\nabla^{\gamma}A_{ 
\alpha} - 
A^{\alpha}A^{\beta}\nabla_{\mu}A_{\alpha}\nabla_{\nu}A_{\beta})     \nonumber \\
&&  -(g_{\mu\nu}A^{
\alpha}\nabla^{\beta}A_{\alpha}\nabla_{\gamma}\nabla_{\beta}A^{
\gamma} 
-A^{\alpha}A^{\beta}g_{\mu\nu}R_{\beta\gamma}\nabla^{\gamma}A_{\alpha})] 
 \nonumber \\ 
&&+36  (g_{\mu\nu}\nabla^{\beta}\nabla_{\sigma}A^{
\sigma}\nabla_{\gamma}\nabla_{
\beta 
}A^{\gamma}     \!  -\! A^{\alpha}g_{\mu\nu}R_{\alpha\gamma
}\nabla^{\gamma}\nabla_{\beta}A^{\beta})
       \nonumber \\
&&     
+72A^{\alpha}\nabla_{\beta}\nabla_{(\nu}A^{\beta}\nabla_{\mu)}A_{\alpha} 
-\nabla_{\mu}R\nabla_{\nu}R 
       \nonumber \\
&& +6A^{\alpha}R_{\alpha(\nu}\nabla_{\mu)}R 
 -6\nabla_{\alpha}\nabla_{(\nu}A^{\alpha}\nabla_{
\mu)}R     \nonumber \\
&&
+12A^{\alpha}\nabla_{(\mu}R\nabla_{\nu)}A_{\alpha}  
-72A^{\alpha}A^{\beta}R_{\beta(\mu}\nabla_{\nu)}A_{\alpha}
\Big\}
f_{\tilde{R}\tilde{R} 
\tilde{R}} ,
\end{eqnarray} 
 where a subscript denotes the derivative of   function 
$f$ in terms of that argument.
Additionally, variation of the action with respect to the Weyl field gives
\begin{eqnarray}
     &&   \nabla_{\alpha}F^{\alpha\mu} 
        +2A^{\mu}f_{\mathcal{A}} 
        -12A^{\mu}(1+f_{\tilde{R}})
        -12A^{\alpha}\nabla^{\mu}A_{\alpha}f_{\tilde{R}\mathcal{A}}\nonumber \\
        &&+36A^{\alpha}R^{\mu}{}_{\alpha}f_{\tilde{R}\tilde{R}}
        +72A^{\alpha}\nabla^{\mu}A_{\alpha}f_{\tilde{R}\tilde{R}}
     \nonumber \\  
&&-6\nabla^{\mu}Rf_{\tilde{R}\tilde{R}}-36\nabla_{\alpha}\nabla^{\mu}A^{\alpha}
f_ {\tilde{R}\tilde{R}}
        =0.        \label{Aeqgen}
\end{eqnarray}
Once again,    imposing the matter conservation equation  $\nabla_{\mu} 
T^{\mu}{}_{\nu} = 0$, equation  (\ref{Weylfieldeqs}) implies 
$ \nabla_{\mu} K^{\mu}{}_{\nu} = 0$ too.

The above equations are the most general equations of modified gravity from 
Weyl connection. In this case one has an  
extra vector degree  of freedom ({three propagating modes}) comparing to 
general relativity, namely the Weyl 
field, 
while another extra scalar degree of freedom, {the scalaron}, arises as 
usual from the 
$f(\tilde{R})$ part, and thus 
from 
the higher-than-linear terms of the Ricci scalar. Hence, 
the higher derivative 
terms that appear in the metric field equations are not problematic, since they 
signal the presence of the extra degree of freedom and can be eliminated 
through a conformal transformation. On the other hand, note that the Weyl field 
equation of motion is second-order. Thus, the class of theory at hand is also 
free from Ostrogradsky ghosts. {Finally, note that this general class of 
theories does not anymore fall with the generalised Proca theories 
\cite{Heisenberg:2014rta}, due to the presence of the nonlinear-in-$R$ terms.
 }

Lastly, let us make a comment on gauge invariance. As mentioned above, the 
initial introduction of the Weyl field was made for gauge invariance reasons. 
In general, by upgrading it to a dynamical field and introducing   
potentials and couplings would not respect this invariance, unless the various 
terms are carefully chosen in order to  be gauge invariant (under the 
known gauge transformations of the Weyl field), or using other 
approaches such as the introduction of  the
 St\"uckelberg mechanism. Nevertheless, even in the general cases where no 
care is devoted to the potentials and couplings choices, the presented theories 
are justifiable, considered  in the effective field theory (EFT) framework 
where any term consistent with the spacetime symmetries  and the desired 
field content can appear in the action (gauge invariance is not a strict 
requirement in effective theories, since symmetry-breaking terms can 
emerge as part of low-energy or effective descriptions of more fundamental 
dynamics by integrating out degrees of freedom).

 \section{Cosmology}
 \label{Cosmology}
 
 In the previous section we constructed gravitational theories on Weyl 
geometry.  In the present section we are interested in  applying them in a 
cosmological framework. To achieve this we consider the homogeneous and 
isotropic flat Friedmann-Lemaître-Robertson-Walker (FLRW) metric,  
namely
\begin{equation}
ds^{2}=-dt^{2}+a(t)^{2} \delta_{ij}dx^{i}dx^{j},
\end{equation}
where $a(t)$ is the scale factor. In order for the matter fluid to respect 
homogeneity 
and isotropy as usual we impose the matter energy-momentum tensor to have the 
form $ T^{\mu}{}_{\nu}=\text{diag}( -\rho_m(t), p_m(t), p_m(t), p_m(t))$, where 
the matter energy density and pressure depend only on time. Similarly, in order 
for $A_\mu$ to  respect the same symmetries we impose the simplest ansatz    
$A_{\mu} = (A_0(t),0,0,0)$.  Note that under this ansatz the field 
strength tensor $F_{\mu\nu}$ of the Weyl gauge field vanishes identically. 

\subsection{General Cases}

We proceed by  inserting this cosmological setup in the classes of theories 
presented in the previous section. Since Class I coincides with general 
relativity, we proceed to Class II.

\subsubsection{Class II}

Let us first study the case of action (\ref{fullaction}).  
The    equation of motion for the Weyl field (\ref{Weylconnectioneqs}) gives
either $A_0=0$ (in which case we recover general relativity), or $
    f'(\mathcal{A})= 6$, which then leads to
\begin{eqnarray}
 f(\mathcal{A}) = 6 
\mathcal{A} + C,
\end{eqnarray}
with   $\mathcal{A}= -A_0^2(t)$ and   $C$ being the integration constant.
 Hence, inserting the above into the     field equations (\ref{Weylfieldeqs}) 
with  (\ref{KtensorclassII})
we obtain the 
Friedmann equations  
\begin{eqnarray}
\label{Fr1a}
H^{2}+\frac{k}{a^2}&=&\frac{8\pi G}{3}\rho_m+ \frac{\Lambda_{eff}}{3}\\
\label{Fr2s}
\dot{H}+H^2&=&- \frac{4\pi G}{3} (\rho_m + 3p_m)+  \frac{\Lambda_{eff}}{3} ,
\end{eqnarray}
where   $H=\dot{a}/a$ is the Hubble function, with dots 
denoting derivatives 
with respect to $t$, and where we have defined
  $\Lambda_{eff}\equiv-\frac{C}{2}$.   
  
Interestingly enough, in this simple case  of modified gravity from Weyl 
connection and geometry we recover
$\Lambda$CDM cosmology,  with 
an effective cosmological constant of geometrical origin,
namely that arises from the richer structure of Weyl geometry. This is one of 
the main results of the present work.

\subsubsection{Class III}
 
Although obtaining an effective cosmological constant is already a success of 
the construction, we proceed to richer structures, that could lead to 
richer phenomenology, too. Hence, we examine action (\ref{actionclassIII}).
In this case,  the equation of motion for the Weyl   field  
(\ref{Weylconnectioneqs22}) gives  

\begin{equation} \label{EOMA}
   A_0\left[ A_0h(\mathcal{A})(4A_0-3H) - 
2A_0^4h'(\mathcal{A})+12-2f'(\mathcal{A}) \right] =0, 
\end{equation} 
 while the       field equations (\ref{Weylfieldeqs}) 
with  (\ref{KtensorclassIII}) yield the two Friedman equations, namely 
\begin{eqnarray}
&&  \!\!\!\!\!\!\!\!\ \!\!\!\!\!\!\!\!\!     H^2 = \frac{8\pi 
G}{3}\rho_m +A_0^2 - \frac{1}{6}  
\left[3A_0^3  
h(\mathcal{A})(H-A_0 )+f(\mathcal{A})\right. \nonumber \\
  &   &
  \ \ \ \ \  \ \ \ \ \  \ \ \  \ \ \  \ \ \  \ \ \  \ \ \,  
  \left. +2A_0^6h'(\mathcal{A})+2A_0^2f'(\mathcal{A}) 
\right] ,
\end{eqnarray}
 and 
\begin{eqnarray}
&&  \!\!\!\!\!\!  \!\!\!\!\!\!  \!\!\!\!\!\!  \!\!\!\!\!\!  \!\!\!\!\!
        H^2+\dot{H} =  -\frac{4\pi G}{3}(\rho_m+3p_m)-2A_0^2 \nonumber \\
        &&-\frac{1}{4}A_0^2\left[2A_0^2
+\dot{A}_0 -A_0 H \right] h(\mathcal{A})  \nonumber \\
        &&+ \frac{1}{6} \left[  -f(\mathcal{A}) 
+ A_0^6h'(\mathcal{A})  + A_0^2f' (\mathcal{A}) \right]   .    
\end{eqnarray}
As mentioned in the previous section, these equations are second-ordered, and 
thus ghost free.
We can re-write the above Friedmann equations in the standard form 
\begin{eqnarray}\
\label{Fr1}
&&    H^2 = \frac{8\pi G}{3}(\rho_m+\rho_{DE})\\
&&   H^2+\dot{H} = -\frac{4\pi G}{3}(\rho_m+3p_m +\rho_{DE}+ 3p_{DE})
\label{Fr2}
\end{eqnarray}
by introducing an effective dark energy sector of geometrical origin, with 
energy density and pressure respectively given by  
 \begin{eqnarray}
&& 
\!\!\!\!\!\!\!\!\!\!\!\!\!\!
\rho_{DE}= \frac{3}{8\pi G}\Big\{   A_0^2 - \frac{1}{6}  
\Big[3A_0^3  
h(\mathcal{A})(H-A_0 )+f(\mathcal{A})   \nonumber \\
  &   &
  \ \ \ \ \  \ \ \ \ \  \ \  \  \ \ \  \ \ \  \ \ \,  
    +2A_0^6h'(\mathcal{A})+2A_0^2f'(\mathcal{A}) 
\Big] \Big\},
\label{rhode}
\end{eqnarray} 
and
  \begin{equation}
 p_{DE}=  \frac{1}{16\pi G}
 \left[A_0^4h(\mathcal{A}) + f(\mathcal{A}) 
+6A_0^2+A_0^2h(\mathcal{A})\dot{A}_0\right].
\label{pde}
\end{equation} 
Furthermore, we can define the effective 
dark-energy equation-of-state parameter as
\begin{eqnarray}
w_{DE}\equiv \frac{p_{DE}}{\rho_{DE}}\,.
\label{wdedef}
\end{eqnarray}
Finally, from (\ref{Fr1}),(\ref{Fr2}), and assuming that the matter sector is 
conserved, i.e.
\begin{eqnarray}
\dot{\rho}_{m}+3H(\rho_{m}+p_{m})=0\,,
\end{eqnarray}
we obtain 
\begin{eqnarray}
\dot{\rho}_{DE}+3H(\rho_{DE}+p_{DE})=0\,,
\end{eqnarray}
which implies that the effective dark energy sector is conserved, which 
according to (\ref{ConservKIII})  was 
expected. 

In summary,   
the richer structure of Weyl geometry gives rise to a dynamical effective dark 
energy. Note that  if we set $f=h=0$ we recover 
standard general relativity, since in this case (\ref{EOMA}) 
gives $A_{\mu}=0$ too and thus the Weyl Ricci scalar $\Tilde{R}$ becomes the 
usual Levi-Civita Ricci scalar 
$R$.

\subsubsection{Class IV}

In the case of the general class of theories determined by   action 
(\ref{GeneralAction}),  the field equations  (\ref{Weylfieldeqs}) with 
(\ref{GeneralK}) provide the two 
Friedmann equations, namely 
 \begin{equation}
    \begin{aligned}
        H^2 =& \frac{8\pi G}{3}\rho_m +\frac{1}{3}K^0{}_0
    \end{aligned}
\end{equation}
and  
\begin{equation}
    \begin{aligned}
        \dot{H} + H^2 = -\frac{4 \pi G }{3} (\rho_m+3p_m) - \frac{1}{6}K^0{}_0 + 
\frac{1}{2}K^1{}_1,
    \end{aligned}
\end{equation}
where  
\begin{eqnarray}
 &&  \!\!\!\!\!\!\!\!\!\!\!\!\!\!\!\!    K^0{}_0 = 
3(\dot{H}+H^2)f_{\tilde{R}} 
        +3A_0H(-3f_{\tilde{R}}+2\dot{A}_0f_{\tilde{R}\mathcal{A}})
                \nonumber
        \\ \ \ \,
        &&-\frac{1}{2}f 
        -3\dot{A}_0f_{\tilde{R}} 
        -A_0^2(-3+f_A-6f_{\tilde{R}}+6\dot{A}_0f_{\tilde{R}\mathcal{A}})  
\nonumber      
\\ \ \ \,
        &&+18f_{\tilde{R}\tilde{R}}(A_0-H) (4H\dot{H}+\ddot{H}+ 
2\dot{A}_0A_0 \nonumber      
\\ \ \ \, 
&& \ \ \ \ \ \ \ \ \ \ \ \  \ \ \ \ \ \ \ \ 
\  \ \ \ -\ddot{A}_0-3H\dot{A}_0-3A_0\dot{H})
\end{eqnarray}
and
\begin{eqnarray}
&& 
\!\!\!\!\!\!\!\!
K^1{}_1 = 3H^2f_{\tilde{R}} 
        +\dot{H}f_{\tilde{R}} 
        -\frac{1}{2}f
        -3\dot{A}_0f_{\tilde{R}} 
        +2\dot{A}_0^2f_{\tilde{R}\mathcal{A}} \nonumber \\
        &&-6f_{\tilde{R}\tilde{R}} \!\left[   (\dot{H}+H^2)^2 -H^4 
         - 5H\ddot{A}_0
        + 6H\ddot{H}+2\dot{A}_0^2
        \right.\nonumber  \\
        &&\left. \ \ \ \ \ \ \ \ 
                -{\dddot{A_0}}
    -6\dot{A}_0(\dot{H}+H^2)+6H^2\dot{H}+3\dot{H}^2
    +\dddot{H}\right] \nonumber \\   
    &&
+72A_0f_{\tilde{R}\tilde{R}\tilde{R}}(2\dot{A}_0-3\dot{H})
(\ddot{A}_0+3H\dot{A}_0-4H\dot{H}-\ddot{H}
)\nonumber 
\\
&&-36f_{\tilde{R}\tilde{R}\tilde{R}}(H^3\!+\!3H\dot{A}_0\!-\!\dot{H}
\!-\!H^2\!+\!\ddot {A }
_0\!-\!3\dot {H}H\!-\!\ddot{H})^2 \nonumber \\
&&
-3A_0^2
    +6\dot{A}_0A^2_0f_{\tilde{R}\mathcal{A}}
    -36\dot{A}_0A^2_0f_{\tilde{R}\tilde{R}}
    -4\dot{A}_0^2A_0^2f_{\tilde{R}\mathcal{A}\mathcal{A}}\nonumber  \\
  &  &+48\dot{A}_0^2A_0^2f_{\tilde{R}\tilde{R}\mathcal{A}}   
+18H^2(-3f_{\tilde{R}\tilde{R}}+4\dot{A}_0f_{\tilde{R}\tilde{R}\mathcal{A}}
)\nonumber 
\\
   & &-36f_{\tilde{R}\tilde{R}\tilde{R}}(2\dot{A}_0-3\dot{H})^2
   \nonumber  \\
&&+2A_0\ddot{A}_0(f_{\tilde{R}A}+3f_{\tilde{R}\tilde{R}}-12\dot{A}_0f_{\tilde{R}
\tilde{R}\mathcal{A}}) \nonumber 
\\
&&+12H^3A_0(3f_{\tilde{R}\tilde{R}\tilde{R}} - 4\dot{A}_0 
f_{\tilde{R}\tilde{R}\mathcal{A}} 
)\nonumber  \\
&&+12HA_0(\dot{H}+H^2)(-3f_{\tilde{R}\tilde{R}}+2\dot{A}_0f_{\tilde{R}\tilde{R}
\mathcal{A}
})\nonumber 
\\
&&+A_0H(-9f_{\tilde{R}} +4\dot{A}_0f_{\tilde{R}\mathcal{A}} + 
30\dot{A}_0f_{\tilde{R}\tilde{R}} )\nonumber  \\
&&
+24A_0\dot{A}_0f_{\tilde{R}\tilde{R}\mathcal{A}}(-3H\dot{A}_0 +H^3+3H\dot{H} 
+\ddot{H} 
).
\end{eqnarray}
Finally the equation of motion for $A_{\mu}$ reads 
\begin{eqnarray}     
&&
18\!
\left(2H^3-6A_0H^2-   3H\dot{A}_0-3A_0\dot{H} +2H\dot{H}+\ddot{H} 
\right)f_{\tilde{R}\tilde{R}}  \nonumber  \\
  &      &+A_0\left[ 
f_{\mathcal{A}}-6f_{\tilde{R}}+6\dot{A}_0(f_{\tilde{R}\mathcal{A}}-6f_{\tilde
{R}\tilde{R}})
\right.
 \nonumber  \\
  &      &\left. \ \ \ \ \ \ \ \, 
+18\ddot{A}_0f_{\tilde{R}\tilde{R}} -6\right] = 0 .
\label{eomAcaseIV}
\end{eqnarray}
 Similarly to the previous class of theories,  we can re-write the above 
Friedmann equations in the standard form (\ref{Fr1}),(\ref{Fr2})
by introducing an effective dark energy sector  with 
energy density and pressure 
 \begin{eqnarray}
&& \rho_{DE}=  \frac{ K^0{}_0}{8\pi G} 
\label{rhodegen}\\
&& 
 p_{DE}=  -
 \frac{  K^1{}_1}{8\pi G} .
\label{pdegen}
\end{eqnarray} 
Moreover,  the effective 
dark-energy equation-of-state parameter is $
w_{DE}\equiv \frac{p_{DE}}{\rho_{DE}}$,
while from the two Friedman equations we obtain 
$
\dot{\rho}_{DE}+3H(\rho_{DE}+p_{DE})=0$  and thus 
 the effective dark energy sector is conserved, as expected from $ \nabla_{\mu} 
K^{\mu}{}_{\nu} = 0$.
Lastly, when $f=0$ the theory recovers 
standard general relativity, since in this case (\ref{eomAcaseIV}) leads to 
$A_{\mu}=0$, and therefore the Weyl Ricci scalar $\Tilde{R}$ becomes the usual 
Levi-Civita Ricci scalar $R$. 
 
\subsection{Specific example}

The general classes of theories  from Weyl 
connection and geometry presented above can have a huge variety of 
cosmological applications.  Nevertheless, for completeness, we close this 
first work    by examining a specific example. Since Class II 
recovers $\Lambda$CDM cosmology, we proceed to Class III, and we consider the 
simple model where 
 \begin{eqnarray}
&& h(\mathcal{A})=\frac{\beta}{\mathcal{A}} 
\nonumber\\
&&  f(\mathcal{A})=\gamma,\label{example1}
\end{eqnarray} 
with $\beta$ and $\gamma$ constant parameters.
 In this case, and recalling that $\mathcal{A}=-A_0^2$ the  Weyl 
field equation (\ref{EOMA}) gives simply 
\begin{equation} \label{Asolexample}
  A_0= \frac{3\beta H}{2(\beta-6)}
\end{equation}
for $\beta\neq6$ and $A_0=7H/2$ for $\beta=6$.
Hence, inserting the above  ans\"atze for the $h(\mathcal{A})$ and 
$f(\mathcal{A})$ functions, alongside the Weyl field solution 
(\ref{Asolexample}), into
the effective dark energy 
density and pressure (\ref{rhode}),(\ref{pde}), we obtain
 \begin{equation}
 \rho_{DE}
    =\frac{1}{8\pi G} \left[ -\frac{\gamma}{2}+ \frac{9 \beta^2 
H^2}{8(\beta-6)}
\right] 
\label{rhodeex}
\end{equation} 
  \begin{eqnarray}
&&\!\!\!\!\!\!\!\!
 p_{DE}=  \frac{1}{8\pi G}\left[ 
\frac{\gamma}{2}-\frac{3\beta^2}{8(\beta-6)}\left( 3H^2+2 \dot{H} \right) 
\right],
\label{pdeex}
\end{eqnarray} 
while  the
dark-energy equation-of-state parameter is written as \footnote{Note that in 
this specific example we obtain a dark energy density that lies withing the 
running vacuum theories \cite{Sola:2016jky}, nevertheless the corresponding 
dark-energy equation-of-state parameter is not $-1$, and thus the scenario at 
hand is   different.}
 \begin{equation}
  w_{DE}\equiv \frac{p_{DE}}{\rho_{DE}} = \frac{  
\frac{\gamma}{2}-\frac{3\beta^2}{8(\beta-6)}\left( 3H^2+2 \dot{H} \right)  }{  
-\frac{\gamma}{2}+ \frac{9 \beta^2 
H^2}{8(\beta-6)}}.
\label{wdeexample}
 \end{equation}

   \begin{figure}[ht]
    \hspace{-0.7cm}
\includegraphics[scale=0.44]{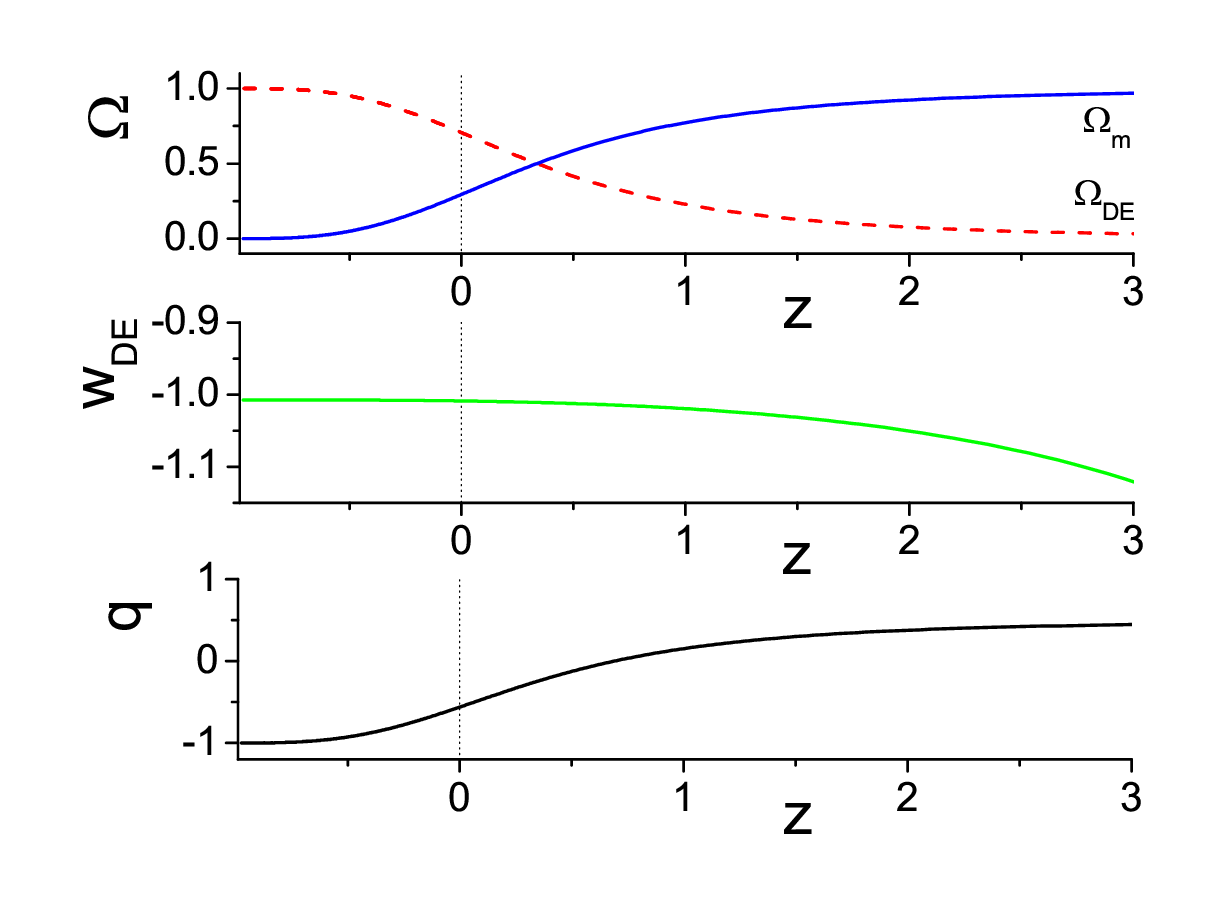}
\caption{
{\it{ Upper graph: The evolution of   the dark energy    
density parameter $\Omega_{DE}$ (blue-solid)  and of the matter density
parameter $\Omega_{m}$ (red-dashed), 
of modified 
gravity with Weyl 
connection, for the specific model (\ref{example1}) within Class III of 
theories, 
as a function of the redshift $z$,  for
 $\beta=0.1$ and $\gamma=-1$ in units 
where $H_0=1$.
 Middle graph: The evolution of the corresponding dark
energy equation-of-state parameter $w_{DE}$ from  (\ref{wdeexample}). Lower 
graph:  The evolution of the 
corresponding   deceleration parameter $q$ from (\ref{decpar}). In all graphs 
we   impose 
$\Omega_{DE}(z=0)\equiv\Omega_{DE0}\approx0.7$  at present, and we 
have added a vertical dotted line denoting 
the current time $z=0$.
}} }
\label{Omegas}
\end{figure}

In order to examine the cosmological evolution  in more detail,  we focus on 
the 
dust-matter case, namely 
we set $p_m=0$. Additionally, we introduce the 
density parameters 
 \begin{eqnarray} \label{FRWomatter}
&&\Omega_m=\frac{8\pi G}{3H^2} \rho_m\\
&& \label{FRWode}
\Omega_{DE}=\frac{8\pi G}{3H^2} \rho_{DE},
 \end{eqnarray} 
 where the subscript ``0" denotes the   value of a quantity at present 
time. Finally, it proves convenient to introduce the 
 deceleration parameter $q $  given as 
  \begin{eqnarray} 
   q\equiv 
-1-\frac{\dot H}{H^2}.
\label{decpar}
  \end{eqnarray}

  As usual,   we use the redshift  $ 1+z=a_0/a$ as the independent variable and 
we set the current scale factor $a_0=1$.
We elaborate  the cosmological equations numerically, imposing   
$\Omega_{DE}(z=0)\equiv\Omega_{DE0}\approx0.7$ and   
$\Omega_m(z=0)\equiv\Omega_{m0}\approx0.3$  as required by 
observational data
\cite{Aghanim:2018eyx}.  In the upper panel of Fig. \ref{Omegas} we depict the 
density parameters 
$\Omega_{DE}(z)$ and $\Omega_{m}(z)=1-\Omega_{DE}(z)$ as a function of the 
redshift. Moreover, in the middle panel we draw the 
corresponding evolution of the
dark-energy equation-of-state parameter $w_{DE}(z)$. Lastly,  in the 
lower panel we present the evolution of the deceleration parameter.
Note that for clarity we have  
extended the 
evolution   to the   future, namely in the region
$z \rightarrow -1$.

 \begin{figure}[ht]
 \hspace{-0.7cm}
\includegraphics[scale=0.46]{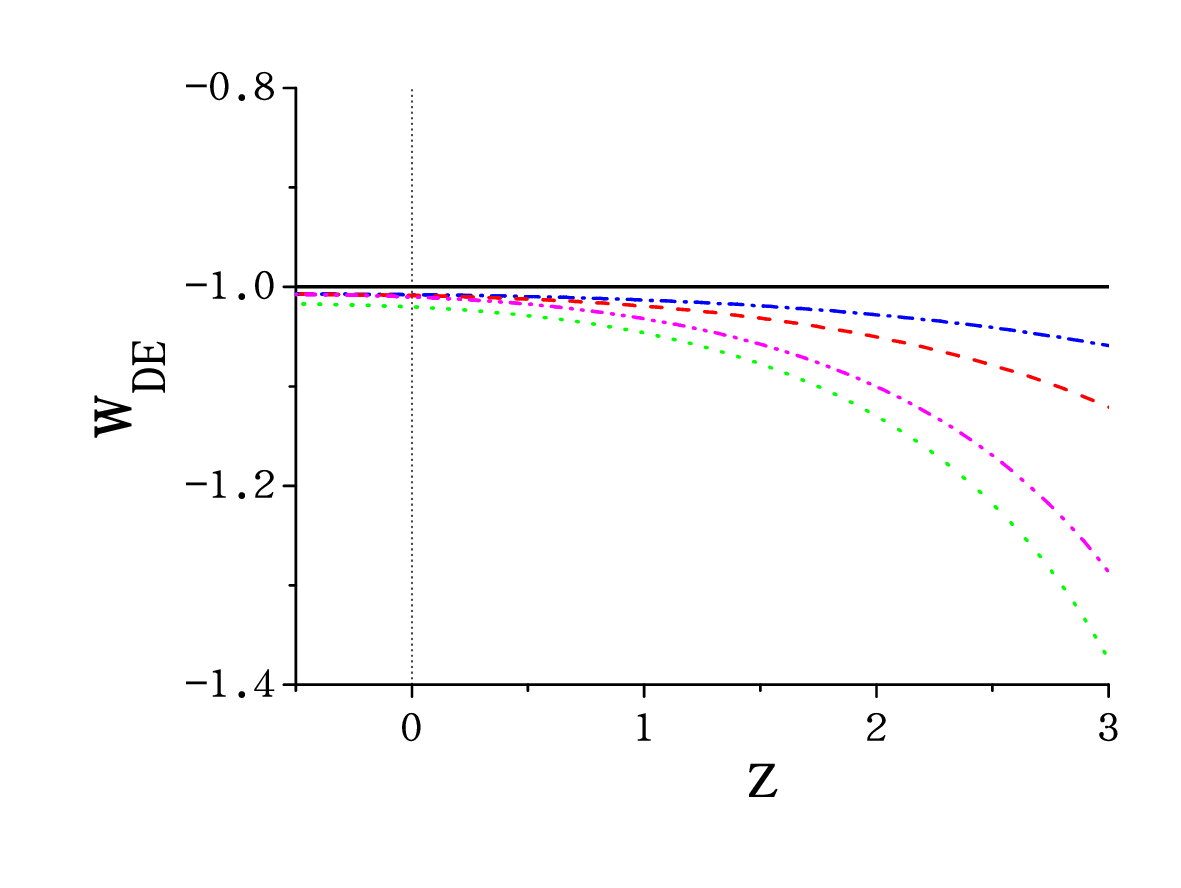}
\caption{
{\it{The evolution of the equation-of-state parameter $w_{DE}$ of modified 
gravity with Weyl 
connection, for the specific model (\ref{example1}) within Class III of 
theories, 
as a function of the redshift $z$,    for 
$\beta=0$,$\gamma=-1$  (black-solid), $\beta=0.1$,$\gamma=-1$ 
(red-dashed), $\beta=0.15$,$\gamma=-1$ (green-dotted)
 $\beta=0.1$,$\gamma=-2$  (blue-dashed-dotted) and  $\beta=0.1$,$\gamma=-0.5$ 
(magenta-dashed-dot-dotted), in units 
where $H_0=1$.
In all graphs we have imposed 
$\Omega_{DE}(z=0)\equiv\Omega_{DE0}\approx0.7$ at present,  and we 
have added a vertical dotted line denoting 
the current time $z=0$.}} }
\label{wplot}
\end{figure}

As we observe, in the scenario at hand we obtain the standard thermal 
history of the universe, i.e. the sequence of matter and dark energy eras, 
while in the future the universe is led asymptotically to the complete 
dark-energy domination. Additionally,  we can see that the transition 
from deceleration to acceleration takes place  at $z\approx 0.6$ in 
agreement with observations. Finally, concerning  $w_{DE}$, we can see that 
its current value    is around $-1$ in agreement with observations, 
nevertheless as described above, it has a dynamic behavior.

Since the effective dark energy exhibits a dynamical nature, it would be 
interesting to examine the behavior of $w_{DE}$ according to the model 
parameters. Thus, in Fig. \ref{wplot} 
we 
draw $w_{DE}(z)$ for various values of  $\beta$ and $\gamma$. As we can see, 
for $\beta=0$ the scenario recovers  $\Lambda$CDM model, while 
for increasing $\beta$  the present value $w_{DE}(z=0)$  tends to 
lower 
values, and on the other hand for  increasing $\gamma$ the scenario comes 
closer to  $\Lambda$CDM paradigm. Finally, we mention that in this specific 
example 
 $w_{DE}$  lies in the phantom 
regime, 
since according to relation (\ref{wdeexample}) this is allowed in the model at 
hand, which is an additional advantage. In summary, through this sample example 
we showed that  modified gravity with Weyl connection can lead to interesting 
cosmological phenomenology.

 \section{Conclusions}
 \label{Conclusions}

In this manuscript we used  Weyl connection and Weyl geometry in order to 
construct novel modified gravitational theories. In particular, it is known 
that in Weyl geometry one uses the Weyl-invariant connection which differs from 
the Levi-Civita connection by terms of the extra Weyl   field. Hence, one 
can construct the corresponding Riemann tensor and Ricci scalar and use it as a 
building block for modified theories of gravity. 

As we showed, in the simplest 
case where one uses only the  Weyl-connection  Ricci scalar as a Lagrangian the 
theory recovers general relativity and no new information is obtained. However, 
by upgrading the Weyl field to a dynamical field with a general potential 
and/or general couplings  constructed from its trace, leads to new modified 
gravity theories,  where the extra vector degree of freedom comparing to 
general 
relativity is the Weyl field. Additionally, since the  
Weyl-connection   Ricci scalar, differs from the  Levi-Civita Ricci scalar by 
terms up to first derivatives of the Weyl field,  the resulting field equations 
for both the metric and the Weyl field are of second order, and thus  the 
theory is free from Ostrogradsky ghosts.   Finally, we constructed the most 
general theory, namely the  $f(\tilde{R},\cal{A})$ gravity, with the extra  
degrees of freedom being 
the Weyl field and the usual scalaron that is hidden inside the nonlinear 
function of the Ricci scalar. {The subclasses of the theories that are 
linear in the Ricci scalar fall within the generalised Proca theories, however 
in the present case the extra vector field is not added ad hoc but it arises 
from the underlying connection itself. On the other hand, the general 
$f(\tilde{R},\cal{A})$ gravity is more general than the Proca theories, due 
to the additional presence of the scalaron degree of freedom}.

Although one can choose the involved functions in 
order to respect the Weyl gauge invariance, even in the general cases  the 
presented theories are justifiable, considered  in the  
EFT framework where all terms consistent with the spacetime symmetries  and the 
desired field content can appear in the action, since symmetry-breaking terms 
can emerge as part of    effective descriptions by integrating out fundamental 
degrees of freedom.

Applying the above classes of theories at a cosmological framework we showed 
that we acquire extra terms in the Friedmann equations, obtaining an effective 
dark energy sector of geometrical origin. In the simplest class of theories we 
were able to obtain an effective cosmological constant, and thus to recover 
$\Lambda$CDM paradigm. Nevertheless, in more general cases we acquired a 
dynamical dark energy, arising from the dynamics of the Weyl field and the 
metric. Hence, the richer geometrical structure of Weyl connection and 
geometry, when applied at a cosmological framework, gives rise to richer and 
interesting cosmological phenomenology.

In particular, we showed that these theories can reproduce the thermal
 history of the Universe, with the sequence of matter and dark-energy epochs.  
Moreover,  the corresponding dark energy equation-of-state parameter presents a 
rich behavior for the various classes of theories and   can be 
quintessence-like, phantom-like, or experience the phantom-divide crossing. In 
the specific example that we provided for completeness,  the 
deceleration-acceleration transitions  takes place   at $z\approx0.6$ in 
agreement with observations,  before the Universe results to  a complete dark 
energy domination in the far future, while the dark energy equation-of-state 
parameter lies in the phantom regime. Such a feature may act as an advantage in 
alleviating the  Hubble tension, since we know that one of the   late-time 
mechanisms that can increase $H_0$ is the phantom dark energy.

In summary, we saw that the  Weyl connection and geometry can be used as a 
basis for the construction of novel modified  gravity theories. We mention here 
that although some classes of these theories lie effectively within the general 
class of Horndeski and generalized Galileon theories (despite the completely 
different origin) \cite{Horndeski:1974wa,DeFelice:2010nf},
this is not true for the most general cases. It would be both interesting and 
necessary to confront the theories with observational data from Supernovae (SN 
Ia), Baryon Acoustic Oscillations (BAO), Cosmic Microwave Background (CMB), and 
Hubble parameter observations, in order to extract constrains on the viable 
classes of theories and parameter spaces. Additionally, one could perform a 
dynamical-system analysis, in order to reveal the global features of the 
scenarios, independently of the specific initial conditions.   Finally, one 
should  investigate the theories at the perturbative level, since one expects 
novel features due  to the richer connection structure. All these necessary 
studies lie beyond the scope of this first work on the subject, and will be 
performed in separate projects.

\begin{acknowledgments} 
The authors acknowledge the contribution of the LISA CosWG, and of   COST 
Actions   CA21136 ``Addressing observational tensions in cosmology with 
systematics and fundamental physics (CosmoVerse)'', CA21106 ``COSMIC WISPers 
in the Dark Universe: Theory, astrophysics and experiments'', 
and CA23130 ``Bridging high and low energies in search of quantum gravity 
(BridgeQG)''.  

 \textbf{Data Availability Statement} This manuscript has no associated data or the data will not be deposited. [Authors’ comment: Data sharing not applicable to this article as no datasets were generated or analysed during the current study.]
 
 \textbf{Code Availability Statement} This manuscript has no associated
code/software. [Author’s comment: Code/Software sharing not applica-
ble to this article as no code/software was generated or analysed during
the current study.]

\end{acknowledgments}


\begin{thebibliography}{99}




 
\bibitem{Sahni:1999gb}
V.~Sahni and A.~A.~Starobinsky,
Int.\ J.\ Mod.\ Phys. D \textbf{9}, 373-444 (2000)
[\href{\arxiv/astro-ph/9904398}{astro-ph/9904398}].


\bibitem{Peebles:2002gy} 
  P.~J.~E.~Peebles and B.~Ratra,
   Rev.\ Mod.\ Phys.\  {\bf 75}, 559 (2003)
  [\href{\arxiv/astro-ph/0207347}{astro-ph/0207347}].
  
  
 

 
\bibitem{Abdalla:2022yfr}
E.~Abdalla, G.~Franco Abell\'an, A.~Aboubrahim, A.~Agnello, O.~Akarsu, 
Y.~Akrami, G.~Alestas, D.~Aloni, L.~Amendola and L.~A.~Anchordoqui, \textit{et 
al.}
JHEAp \textbf{34}, 49-211 (2022)
[\href{https://arxiv.org/abs/2203.06142}{arXiv:2203.06142}].
  
  
 \bibitem{Copeland:2006wr}
  E.~J.~Copeland, M.~Sami and S.~Tsujikawa,
  Int.\ J.\ Mod.\ Phys.\  D {\bf 15}, 1753 (2006)
 [\href{http://xxx.lanl.gov/abs/hep-th/0603057}
{arXiv:0603057}].
 

 
\bibitem{Cai:2009zp}
  Y.~-F.~Cai, E.~N.~Saridakis, M.~R.~Setare and J.~-Q.~Xia,
  Phys.\ Rept.\  {\bf 493}, 1 (2010)  [\href{http://xxx.lanl.gov/abs/0909.2776}
{{arXiv:0909.2776}}].




\bibitem{CANTATA:2021ktz}
E.~N.~Saridakis \textit{et al.} [CANTATA],
[\href{\arxiv/arXiv:2105.12582}{arXiv:2105.12582}].
   
 
 

\bibitem{Capozziello:2011et}
  S.~Capozziello and M.~De Laurentis,
  Phys.\ Rept.\  {\bf 509}, 167 (2011)
[\href{http://xxx.lanl.gov/abs/1108.6266}
{{arXiv:1108.6266}}].
 

\bibitem{Cai:2015emx} 
  Y.~F.~Cai, S.~Capozziello, M.~De Laurentis and E.~N.~Saridakis,
  Rept.\ Prog.\ Phys.\  {\bf 79}, 106901 (2016).
 [\href{http://xxx.lanl.gov/abs/1511.07586}
 {arXiv:1511.07586}].
 
\bibitem{Nojiri:2017ncd}
S.~Nojiri, S.~D.~Odintsov and V.~K.~Oikonomou,
Phys. Rept. \textbf{692}, 1-104 (2017)
 [\href{http://xxx.lanl.gov/abs/1705.11098}
 {arXiv:1705.11098}].
 

 
\bibitem{Starobinsky:1980te}
  A.~A.~Starobinsky,
  Phys.\ Lett.\ B {\bf 91}, 99 (1980).
  
\bibitem{DeFelice:2010aj}
A.~De Felice and S.~Tsujikawa,
Living Rev. Rel. \textbf{13}, 3 (2010).
 [\href{http://xxx.lanl.gov/abs/1002.4928}
 {arXiv:1002.4928}].


\bibitem{Nojiri:2010wj}
S.~Nojiri and S.~D.~Odintsov,
Phys. Rept. \textbf{505}, 59-144 (2011)
 [\href{http://xxx.lanl.gov/abs/1011.0544}
 {arXiv:1011.0544}].
 



 

\bibitem{Capozziello:2002rd}
S.~Capozziello,
Int. J. Mod. Phys. D \textbf{11}, 483-492 (2002)
[\href{https://arxiv.org/abs/gr-qc/0201033}{arXiv:0201033}].

 
\bibitem{Nojiri:2005jg}
  S.~Nojiri and S.~D.~Odintsov,
  Phys.\ Lett.\ B {\bf 631}, 1 (2005).

\bibitem{DeFelice:2008wz}
A.~De Felice and S.~Tsujikawa,
Phys. Lett. B \textbf{675}, 1-8 (2009)
[\href{https://arxiv.org/abs/0810.5712}{arXiv:0810.5712}].

\bibitem{DeFelice:2009aj}
A.~De Felice and S.~Tsujikawa,
Phys. Rev. D \textbf{80}, 063516 (2009).

\bibitem{Asimakis:2022mbe}
P.~Asimakis, S.~Basilakos and E.~N.~Saridakis,
Eur. Phys. J. C \textbf{84}, no.2, 207 (2024)
[\href{https://arxiv.org/abs/2212.12494}{arXiv:2212.12494}].
   
\bibitem{Lovelock:1971yv}
  D.~Lovelock,
  J.\ Math.\ Phys.\  {\bf 12}, 498 (1971).
  
\bibitem{Deruelle:1989fj}
N.~Deruelle and L.~Farina-Busto,
Phys. Rev. D \textbf{41}, 3696 (1990).

 
   
 
\bibitem{Linder:2010py}
E.~V.~Linder,
Phys. Rev. D \textbf{81}, 127301 (2010)
[erratum: Phys. Rev. D \textbf{82}, 109902 (2010)]
[\href{https://arxiv.org/abs/1005.3039}{arXiv:1005.3039}].
\bibitem{Chen:2010va}
S.~H.~Chen, J.~B.~Dent, S.~Dutta and E.~N.~Saridakis,
Phys. Rev. D \textbf{83}, 023508 (2011)
[\href{https://arxiv.org/abs/1008.1250}{arXiv:1008.1250}].

\bibitem{Kofinas:2014owa}
G.~Kofinas and E.~N.~Saridakis,
Phys. Rev. D \textbf{90}, 084044 (2014).
[\href{https://arxiv.org/abs/1404.2249}{arXiv:1404.2249}].

\bibitem{Kofinas:2014daa}
G.~Kofinas and E.~N.~Saridakis,
Phys. Rev. D \textbf{90}, 084045 (2014).
[\href{https://arxiv.org/abs/1408.0107}{arXiv:1408.0107}].

\bibitem{Bahamonde:2015zma}
S.~Bahamonde, C.~G.~B\"ohmer and M.~Wright,
Phys. Rev. D \textbf{92}, no.10, 104042 (2015).
[\href{https://arxiv.org/abs/1508.05120}{arXiv:1508.05120}].

\bibitem{Bahamonde:2016grb}
S.~Bahamonde and S.~Capozziello,
Eur. Phys. J. C \textbf{77}, no.2, 107 (2017)
[\href{https://arxiv.org/abs/1612.01299}{arXiv:1612.01299}].

\bibitem{BeltranJimenez:2017tkd}
J.~Beltr\'an Jim\'enez, L.~Heisenberg and T.~Koivisto,
Phys. Rev. D \textbf{98}, no.4, 044048 (2018).
[\href{https://arxiv.org/abs/1710.03116}{arXiv:1710.03116}].
   
   
\bibitem{Heisenberg:2023lru}
L.~Heisenberg,
[\href{https://arxiv.org/abs/2309.15958}{arXiv:2309.15958}]

\bibitem{De:2023xua}
A.~De, T.~H.~Loo and E.~N.~Saridakis,
JCAP \textbf{03}, 050 (2024)
[\href{https://arxiv.org/abs/2308.00652}{arXiv:2308.00652}].



 
\bibitem{Weyl:1918ib}
H.~Weyl,
Sitzungsber. Preuss. Akad. Wiss. Berlin (Math. Phys. ) \textbf{1918}, 465 
(1918).

 
\bibitem{Weylbook}
 Eduardo Garcia-Rio, Peter Gilkey , Stana Nikcevic, Ramon Vazquez-Lorenzo
{\it{Applications of Affine and Weyl Geometry}},
Springer, (2013).
 
\bibitem{Mannheim:1988dj}
P.~D.~Mannheim and D.~Kazanas,
Astrophys. J. \textbf{342}, 635-638 (1989).
 
 
\bibitem{Zee:1981ff}
A.~Zee,
Phys. Lett. B \textbf{109}, 183-186 (1982)
 


\bibitem{Zee:1983mj}
A.~Zee,
Annals Phys. \textbf{151}, 431 (1983)
 
\bibitem{Kazanas:1988qa}
D.~Kazanas and P.~D.~Mannheim,
Astrophys. J. Suppl. \textbf{76}, 431-453 (1991)





\bibitem{Sola:1988nz}
J.~Sola,
Phys. Lett. B \textbf{228}, 317-324 (1989)



\bibitem{Mannheim:1990ya}
P.~D.~Mannheim and D.~Kazanas,
Phys. Rev. D \textbf{44}, 417-423 (1991).
  
\bibitem{La:1991nk}
D.~La,
Phys. Rev. D \textbf{44}, 1680-1684 (1991)



\bibitem{Elizondo:1994vh}
D.~Elizondo and G.~Yepes,
Astrophys. J. \textbf{428}, 17-20 (1994)
[\href{https://arxiv.org/abs/astro-ph/9312064}{arXiv:9312064}].



\bibitem{Bronnikov:1997gj}
K.~A.~Bronnikov and J.~C.~Fabris,
Class. Quant. Grav. \textbf{14}, 831-842 (1997)
[\href{https://arxiv.org/abs/gr-qc/9603037}{arXiv:9603037}].

 

\bibitem{Edery:1997hu}
A.~Edery and M.~B.~Paranjape,
Phys. Rev. D \textbf{58}, 024011 (1998)
[\href{https://arxiv.org/abs/astro-ph/9708233}{arXiv:9708233}].



 
\bibitem{Klemm:1998kf}
D.~Klemm,
Class. Quant. Grav. \textbf{15}, 3195-3201 (1998)
[\href{https://arxiv.org/abs/gr-qc/9808051}{arXiv:9808051}].




\bibitem{Boulanger:2001he}
N.~Boulanger and M.~Henneaux,
Annalen Phys. \textbf{10}, 935-964 (2001)
[\href{https://arxiv.org/abs/hep-th/0106065}{arXiv:0106065}].


\bibitem{Pireaux:2004xb}
S.~Pireaux,
Class. Quant. Grav. \textbf{21}, 4317-4334 (2004)
[\href{https://arxiv.org/abs/gr-qc/0408024}{arXiv:0408024}].



\bibitem{Pireaux:2004id}
S.~Pireaux,
Class. Quant. Grav. \textbf{21}, 1897-1913 (2004)
[\href{https://arxiv.org/abs/gr-qc/0403071}{arXiv:0403071}].




\bibitem{Flanagan:2006ra}
E.~E.~Flanagan,
Phys. Rev. D \textbf{74}, 023002 (2006)
[\href{https://arxiv.org/abs/astro-ph/0605504}{arXiv:0605504}].

 
 
\bibitem{Edery:2006hg}
A.~Edery, L.~Fabbri and M.~B.~Paranjape,
Class. Quant. Grav. \textbf{23}, 6409-6423 (2006)
[\href{https://arxiv.org/abs/hep-th/0603131}{arXiv:0603131}].


 
 
\bibitem{Lobo:2008zu}
F.~S.~N.~Lobo,
Class. Quant. Grav. \textbf{25}, 175006 (2008)
[\href{https://arxiv.org/abs/0801.4401}{arXiv:0801.4401}].




\bibitem{Sultana:2010zz}
J.~Sultana and D.~Kazanas,
Phys. Rev. D \textbf{81}, 127502 (2010)



\bibitem{Percacci:2011uf}
R.~Percacci,
New J. Phys. \textbf{13}, 125013 (2011)
[\href{https://arxiv.org/abs/1110.6758}{arXiv:1110.6758}].


\bibitem{Dengiz:2011ig}
S.~Dengiz and B.~Tekin,
Phys. Rev. D \textbf{84}, 024033 (2011)
[\href{https://arxiv.org/abs/1104.0601}{arXiv:1104.0601}].


 


\bibitem{Tanhayi:2012nn}
M.~R.~Tanhayi, S.~Dengiz and B.~Tekin,
Phys. Rev. D \textbf{85}, 064016 (2012)
[\href{https://arxiv.org/abs/1201.5068}{arXiv:1201.5068}].


\bibitem{Sultana:2012qp}
J.~Sultana, D.~Kazanas and J.~Levi Said,
Phys. Rev. D \textbf{86}, 084008 (2012)
[\href{https://arxiv.org/abs/1910.06118}{arXiv:1910.06118}].

\bibitem{Deruelle:2012xv}
N.~Deruelle, M.~Sasaki, Y.~Sendouda and A.~Youssef,
JHEP \textbf{09}, 009 (2012)
[\href{https://arxiv.org/abs/1202.3131}{arXiv:1202.3131}].


\bibitem{Said:2012xt}
J.~L.~Said, J.~Sultana and K.~Z.~Adami,
Phys. Rev. D \textbf{85}, 104054 (2012)
[\href{https://arxiv.org/abs/1201.0860}{arXiv:1201.0860}].



\bibitem{Cattani:2013dla}
C.~Cattani, M.~Scalia, E.~Laserra, I.~Bochicchio and K.~K.~Nandi,
Phys. Rev. D \textbf{87}, no.4, 047503 (2013)
[\href{https://arxiv.org/abs/1303.7438}{arXiv:1303.7438}].



\bibitem{Wheeler:2013ora}
J.~T.~Wheeler,
Phys. Rev. D \textbf{90}, no.2, 025027 (2014)
[\href{https://arxiv.org/abs/1310.0526}{arXiv:1310.0526}].



\bibitem{Quiros:2014hua}
I.~Quiros,
[\href{https://arxiv.org/abs/1401.2643}{arXiv:1401.2643}].

 
\bibitem{Myung:2014jha}
Y.~S.~Myung and T.~Moon,
JCAP \textbf{08}, 061 (2014)
[\href{https://arxiv.org/abs/1406.4367}{arXiv:1406.4367}].

 
 
\bibitem{Cusin:2015rex}
G.~Cusin, S.~Foffa, M.~Maggiore and M.~Mancarella,
Phys. Rev. D \textbf{93}, no.4, 043006 (2016)
[\href{https://arxiv.org/abs/1512.06373}{arXiv:1512.06373}].


\bibitem{Mureika:2016efo}
J.~R.~Mureika and G.~U.~Varieschi,
Can. J. Phys. \textbf{95}, no.12, 1299-1306 (2017)
[\href{https://arxiv.org/abs/1611.00399}{arXiv:1611.00399}].




\bibitem{Oda:2016psn}
I.~Oda,
Eur. Phys. J. C \textbf{77}, no.5, 284 (2017)
[\href{https://arxiv.org/abs/1610.05441}{arXiv:1610.05441}].




\bibitem{Ghilencea:2018dqd}
D.~M.~Ghilencea,
JHEP \textbf{03}, 049 (2019)
[\href{https://arxiv.org/abs/1812.08613}{arXiv:1812.08613}].



\bibitem{Zinhailo:2018ska}
A.~F.~Zinhailo,
Eur. Phys. J. C \textbf{78}, no.12, 992 (2018)
[\href{https://arxiv.org/abs/1809.03913}{arXiv:1809.03913}].

\bibitem{Ghilencea:2018thl}
D.~M.~Ghilencea and H.~M.~Lee,
Phys. Rev. D \textbf{99}, no.11, 115007 (2019)
[\href{https://arxiv.org/abs/1809.09174}{arXiv:1809.09174}].


 

\bibitem{Ghilencea:2019jux}
D.~M.~Ghilencea,
Phys. Rev. D \textbf{101}, no.4, 045010 (2020)
[\href{https://arxiv.org/abs/1904.06596}{arXiv:1904.06596}].
 
 


\bibitem{Ghilencea:2020rxc}
D.~M.~Ghilencea,
Eur. Phys. J. C \textbf{81}, no.6, 510 (2021)
[\href{https://arxiv.org/abs/2007.14733}{arXiv:2007.14733}].


\bibitem{Takizawa:2020dja}
K.~Takizawa, T.~Ono and H.~Asada,
Phys. Rev. D \textbf{102}, no.6, 064060 (2020)
[\href{https://arxiv.org/abs/2006.00682}{arXiv:2006.00682}].

\bibitem{Jawad:2020wlg}
A.~Jawad, Z.~Khan and S.~Rani,
Eur. Phys. J. C \textbf{80}, no.1, 71 (2020)

\bibitem{Geiller:2021vpg}
M.~Geiller, C.~Goeller and C.~Zwikel,
JHEP \textbf{09}, 029 (2021)
[\href{https://arxiv.org/abs/2107.01073}{arXiv:2107.01073}].


\bibitem{Yang:2022icz}
J.~Z.~Yang, S.~Shahidi and T.~Harko,
Eur. Phys. J. C \textbf{82}, no.12, 1171 (2022)
[\href{https://arxiv.org/abs/2212.05542}{arXiv:2212.05542}].

\bibitem{Hell:2023rbf}
A.~Hell, D.~Lust and G.~Zoupanos,
JHEP \textbf{08}, 168 (2023)
[\href{https://arxiv.org/abs/2306.13714}{arXiv:2306.137142}].
 

\bibitem{Karananas:2021gco}
G.~K.~Karananas, M.~Shaposhnikov, A.~Shkerin and S.~Zell,
Phys. Rev. D \textbf{104}, no.12, 124014 (2021)
[\href{https://arxiv.org/abs/2108.05897}{arXiv:2108.05897}].

\bibitem{Roumelioti:2024lvn}
D.~Roumelioti, S.~Stefas and G.~Zoupanos,
Eur. Phys. J. C \textbf{84}, no.6, 577 (2024)
[\href{https://arxiv.org/abs/2403.17511}{arXiv:2403.17511}].
 
\bibitem{Gialamas:2024iyu}
I.~D.~Gialamas and K.~Tamvakis,
[\href{https://arxiv.org/abs/2410.16364}{arXiv:2410.16364}].
 
 
   
   
\bibitem{Haghani:2012bt}
Z.~Haghani, T.~Harko, H.~R.~Sepangi and S.~Shahidi,
JCAP \textbf{10}, 061 (2012)
[\href{https://arxiv.org/abs/1202.1879}{arXiv:1202.1879}].
   
\bibitem{Haghani:2013pea}
Z.~Haghani, T.~Harko, H.~R.~Sepangi and S.~Shahidi,
Phys. Rev. D \textbf{88}, no.4, 044024 (2013)
[\href{https://arxiv.org/abs/1307.2229}{arXiv:1307.2229}].

\bibitem{Haghani:2014zra}
Z.~Haghani, N.~Khosravi and S.~Shahidi,
Class. Quant. Grav. \textbf{32}, no.21, 215016 (2015)
[\href{https://arxiv.org/abs/1410.2412}{arXiv:1410.2412}].




 
\bibitem{Xu:2020yeg}
Y.~Xu, T.~Harko, S.~Shahidi and S.~D.~Liang,
Eur. Phys. J. C \textbf{80}, no.5, 449 (2020)
[\href{https://arxiv.org/abs/2005.04025}{arXiv:2005.04025}].

\bibitem{Yang:2021fjy}
J.~Z.~Yang, S.~Shahidi, T.~Harko and S.~D.~Liang,
Eur. Phys. J. C \textbf{81}, no.2, 111 (2021)
[\href{https://arxiv.org/abs/2101.09956}{arXiv:2101.09956}].


   
\bibitem{Gadbail:2021kgd}
G.~Gadbail, S.~Arora and P.~K.~Sahoo,
Eur. Phys. J. Plus \textbf{136}, no.10, 1040 (2021)
[\href{https://arxiv.org/abs/2108.00374}{arXiv:2108.00374}].
  
  
\bibitem{Harko:2021tav}
T.~Harko, N.~Myrzakulov, R.~Myrzakulov and S.~Shahidi,
Phys. Dark Univ. \textbf{34}, 100886 (2021)
[\href{https://arxiv.org/abs/2110.00358}{arXiv:2110.00358}].

\bibitem{Berezin:2021iof}
V.~A.~Berezin, V.~I.~Dokuchaev, Y.~N.~Eroshenko, Y.~N.~Eroshenko and 
A.~L.~Smirnov,
JCAP \textbf{11}, no.11, 053 (2021)
[\href{https://arxiv.org/abs/2107.06160}{arXiv:2107.06160}].





\bibitem{Gadbail:2021fjf}
G.~N.~Gadbail, S.~Arora and P.~K.~Sahoo,
Eur. Phys. J. C \textbf{81}, no.12, 1088 (2021)
[\href{https://arxiv.org/abs/2110.02726}{arXiv:2110.02726}].
  
  
  
\bibitem{Berezin:2022phu}
V.~A.~Berezin and V.~I.~Dokuchaev,
Class. Quant. Grav. \textbf{40}, no.1, 015006 (2023)
[\href{https://arxiv.org/abs/2207.00057}{arXiv:2207.00057}].


\bibitem{Berezin:2022odj}
V.~A.~Berezin and V.~I.~Dokuchaev,
Int. J. Mod. Phys. A \textbf{37}, no.20n21, 2243005 (2022)
[\href{https://arxiv.org/abs/2203.04257}{arXiv:2203.04257}].
  
\bibitem{Gadbail:2022scf}
G.~N.~Gadbail, S.~Arora, P.~Kumar and P.~K.~Sahoo,
Chin. J. Phys. \textbf{79}, 246-255 (2022)
[\href{https://arxiv.org/abs/2209.04348}{arXiv:2209.04348}].
  
  
\bibitem{Koussour:2023nqr}
M.~Koussour,
Chin. J. Phys. \textbf{83}, 454-466 (2023)
[\href{https://arxiv.org/abs/2303.00665}{arXiv:2303.00665}].

\bibitem{Gadbail:2023enu}
G.~N.~Gadbail, H.~Chaudhary, A.~Bouali and P.~K.~Sahoo,
Nucl. Phys. B \textbf{1009}, 116727 (2024)
[\href{https://arxiv.org/abs/2305.11190}{arXiv:2305.11190}].


 

\bibitem{Bhardwaj:2023lph}
V.~K.~Bhardwaj and P.~Garg,
Can. J. Phys. \textbf{102}, no.8, 441-452 (2024)
[\href{https://arxiv.org/abs/2310.00666}{arXiv:2310.00666}].

  
  
\bibitem{Koussour:2024aez}
M.~Koussour, S.~Myrzakulova and N.~Myrzakulov,
Int. J. Geom. Meth. Mod. Phys. \textbf{21}, no.10, 2440013 (2024)
[\href{https://arxiv.org/abs/2401.04500}{arXiv:2401.04500}].

 

\bibitem{Zhadyranova:2024hbc}
A.~Zhadyranova, M.~Koussour and S.~Bekkhozhayev,
Chin. J. Phys. \textbf{89}, 1483-1492 (2024)
[\href{https://arxiv.org/abs/2406.15409}{arXiv:2406.15409}].


 
  
  
  
  
  
\bibitem{Sakti:2024pze}
M.~F.~A.~R.~Sakti, P.~Burikham and T.~Harko,
Phys. Rev. D \textbf{110}, no.6, 064012 (2024)
[\href{https://arxiv.org/abs/2401.10410}{arXiv:2401.10410}].
 
  
\bibitem{Banados:2024rfy}
M.~Ba\~nados,
equation 
[\href{https://arxiv.org/abs/2402.15675}{arXiv:2402.15675}].
  

\bibitem{Harko:2024fnt}
T.~Harko and S.~Shahidi,
Eur. Phys. J. C \textbf{84}, no.5, 509 (2024)
[\href{https://arxiv.org/abs/2405.04129}{arXiv:2405.04129}].

\bibitem{Bhagat:2024nac}
R.~Bhagat, F.~Tello-Ortiz and B.~Mishra,
analysis and observational validation,''
[\href{https://arxiv.org/abs/2409.17193}{arXiv:2409.17193}].


 






\bibitem{BeltranJimenez:2016wxw}
J.~Beltran Jimenez, L.~Heisenberg and T.~S.~Koivisto,
JCAP \textbf{04}, 046 (2016)
[\href{https://arxiv.org/abs/1602.07287}{arXiv:1602.07287}].


\bibitem{BeltranJimenez:2015pnp}
J.~Beltran Jimenez and T.~S.~Koivisto,
Phys. Lett. B \textbf{756}, 400-404 (2016)
[\href{https://arxiv.org/abs/1509.02476}{arXiv:1509.02476}].
 
 
 
 
 \bibitem{Weyl:1918pdp}
Weyl, H. Reine Infinitesimalgeometrie. Math Z 2, 384–411 (1918). 

\bibitem{Romero:2012hs}
C. Romero(Paraiba U.), J.B. Fonseca-Neto(Paraiba U.), M.L. Pucheu(Paraiba U.),
Class.Quant.Grav. 29 (2012) 155015
[\href{https://arxiv.org/abs/1201.1469}{arXiv:1201.1469}]. 


\bibitem{Wheeler:2018rjb}
J.~T.~Wheeler,
Gen. Rel. Grav. \textbf{50}, no.7, 80 (2018)
[\href{https://arxiv.org/abs/1801.03178}{arXiv:1801.03178}].

 
\bibitem{Scholz:2017pfo}
E.~Scholz,
Einstein Stud. \textbf{14}, 261-360 (2018).



\bibitem{Tomonari:2024ybs}
K.~Tomonari and D.~Blixt,
[\href{\arxiv/arXiv:2410.15056}{arXiv:2410.15056}].

\bibitem{Tomonari:2024lpv}
K.~Tomonari,
[\href{\arxiv/arXiv:2411.11118}{arXiv:2411.11118}].


 
\bibitem{Gialamas:2024uar}
I.~D.~Gialamas and A.~Racioppi,
[\href{\arxiv/arXiv:2412.17738}{arXiv:2412.17738}].


\bibitem{Capozziello:2024lsz}
S.~Capozziello and M.~Shokri,
Phys. Dark Univ. \textbf{46}, 101698 (2024)
[\href{\arxiv/arXiv:2408.17415}{arXiv:2408.17415}].

\bibitem{Barriga:2024hpe}
F.~Barriga, F.~Izaurieta, S.~Lepe, P.~Meza, J.~Mu\~noz, C.~Quinzacara and 
O.~Valdivia,
JCAP \textbf{02}, 003 (2025)
[\href{\arxiv/arXiv:2409.15509}{arXiv:2409.15509}].

 

\bibitem{Ostrogradsky} M.~Ostrogradsky, 
différentielles, 
Mem. Acad. St. Petersbourg
{\bf{VI 4}},  385 (1850).


\bibitem{Heisenberg:2014rta}
L.~Heisenberg,
JCAP \textbf{05}, 015 (2014)
[\href{https://arxiv.org/abs/1402.7026}{arXiv:1402.7026}].
 
 


\bibitem{Sola:2016jky}
J.~Sol\`a, A.~G\'omez-Valent and J.~de Cruz P\'erez,
Astrophys. J. \textbf{836}, no.1, 43 (2017)
[\href{https://arxiv.org/abs/1602.02103}{arXiv:1602.02103}].
 

\bibitem{Aghanim:2018eyx} 
 N.~Aghanim \textit{et al.} [Planck],
Astron. Astrophys. \textbf{641}, A6 (2020)
[erratum: Astron. Astrophys. \textbf{652}, C4 (2021)]
  [\href{\arxiv/arXiv:1807.06209}{arXiv:1807.06209}].
 
 
 

 

\bibitem{Horndeski:1974wa}
G.~W.~Horndeski,
Int. J. Theor. Phys. \textbf{10}, 363-384
(1974).

 
 
 
\bibitem{DeFelice:2010nf}
A.~De Felice and S.~Tsujikawa,
Phys. Rev. D \textbf{84}, 124029 (2011)
[\href{\arxiv/arXiv:1008.4236}{arXiv:1008.4236}].

 
  
  



\end{thebibliography}
\end{document}